\documentclass[aps,prd,floats,amsmath,amssymb,nofootinbib,
superscriptaddress,
notitlepage,
longbibliography,
12pt,
preprintnumbers
]{revtex4-1}

\usepackage[utf8]{inputenc} 
\usepackage{latexsym} 
\usepackage{amsmath}  
\usepackage{amssymb}
\usepackage{hyperref}
\usepackage{color,graphicx,colortbl}
\usepackage{epsfig}
\usepackage{booktabs}
\usepackage{graphicx}
\usepackage{subfigure} %

\usepackage{xcolor}
\hypersetup{
    colorlinks,
    linkcolor={blue!40!black},
    citecolor={cyan!80!black},
    urlcolor={red!50!black}
}
\usepackage{soul}

\begin{document}



\title{Dipoles and chains of solitons in the Friedberg-Lee-Sirlin model}

\author{V\'ictor Jaramillo}
\email[]{jaramillo@ustc.edu.cn}
\affiliation{Department of Astronomy, University of Science and Technology of China, Hefei, Anhui 230026, China}

\author{Shuang-Yong Zhou}
\email[]{zhoushy@ustc.edu.cn}
\affiliation{Interdisciplinary Center for Theoretical Study, University of Science and Technology of China, Hefei, Anhui 230026, China}
\affiliation{Peng Huanwu Center for Fundamental Theory, Hefei, Anhui 230026, China}


\date{\today}


\begin{abstract}
  We construct static axisymmetric multisolitons in the Einstein-Friedberg-Lee-Sirlin model. This theory features a complex scalar field which gains mass through its interaction with a real scalar field that has a non-zero vacuum expectation value. By performing three-dimensional numerical relativity simulations, we identify stable dipolar boson stars in specific regions of the parameter space. Based on the dipole results, non-rotating odd parity chains with and without gravity can also be constructed when the mass of the real scalar field is sufficiently small. However, chains beyond the dipole case are found to be unstable.
\end{abstract}

\preprint{USTC-ICTS/PCFT-24-45}



\maketitle

\section{Introduction}

The investigation of self-gravitating scalar field configurations in 3 dimensional asymptotically flat spacetime has garnered significant interest in recent decades, primarily due to the fundamental role scalar fields play in models of dark matter, inflation and early Universe evolution. Scalar fields can lead to gravitational collapse, forming localized gravitating objects such as boson stars \cite{Kaup:1968zz,Ruffini:1969qy}—compact, stationary and horizonless configurations where the scalar field exhibits harmonic time dependence. Beyond boson stars, flat space solitons, which are spatially localized field configurations, are widely studied across various fields of theoretical physics, including cosmology and quantum field theory. These solitons can be topological, stabilized by a conserved topological number, or non-topological, like Q-balls \cite{Rosen:1968mfz,Friedberg:1976me,Coleman:1985ki}, which arise in field models with continuous global symmetry and exhibit a stationary oscillating internal phase. 

Interestingly, certain gravitating solutions such as some particular models of boson stars, monopoles, and Skyrmions are closely related to their flat space counterparts, representing non-topological solitons or Q-balls, and exhibit intriguing dynamics when coupled with gravity. Q-balls can be realized in two scenarios, the single complex scalar field with a non-renormalizable self-interaction potential found by Coleman in the 80's \cite{Coleman:1985ki} and the earlier two-component model found by Friedberg, Lee and Sirlin which contains a symmetry breaking potential \cite{Friedberg:1976me}. The Friedberg-Lee-Sirlin (FLS) model provides a compelling example of a renormalizable scalar field theory with natural interaction terms, where a complex scalar field gains mass through coupling with a real scalar field that acquires a finite vacuum expectation value via a symmetry-breaking potential. This interaction gives rise to Q-ball solutions, which emerge from the phase rotation of the complex scalar field. When coupled with gravity, these Q-balls manifest as boson stars.

The FLS model has been studied in various contexts. In flat spacetime, it was shown that solutions exist even when the scalar potential vanishes, rendering the real scalar field massless \cite{Loiko:2018mhb}. Additionally, axially symmetric spinning solutions with both even and odd parity were constructed in \cite{Loiko:2018mhb}, and a U(1)-gauged generalization, which generates a dipolar magnetic field, was explored in \cite{Loiko:2019gwk}. In \cite{Zhang}, the superradiance of Q-balls in the FLS model was investigated, demonstrating the potential for amplifying incident waves. When coupled minimally with gravity, the FLS model has been studied in various scenarios. Boson stars in this theory were first constructed in \cite{Kunz:2019sgn}, where spherical and spinning solutions were found to be connected to their flat spacetime counterparts. The spherical solutions were revisited in the U(1)-gauged case in \cite{Kunz:2021mbm} and further studied in \cite{deSa:2024dhj} with a focus on their astrophysical signatures. In \cite{Herdeiro:2023lze}, a vector analogue of the FLS model was introduced to add self-interactions to the Proca model, avoiding hyperbolicity issues. Remarkably, the Einstein-FLS model also supports black hole solutions, as explored in \cite{Kunz:2021mbm,Kunz:2023qfg,Kunz:2024uux}.

Einstein’s gravity, minimally coupled to free scalar fields, gives rise to families of static solutions beyond the spherical case, exhibiting rich structures. These are known as multipolar boson stars \cite{Herdeiro:2020kvf}. The simplest of these non-spherical configurations is the dipolar boson star \cite{Yoshida:1997nd,Cunha:2022tvk}, a two-center soliton in equilibrium due to the balance between gravitational attraction and repulsion caused by the phase difference in the odd-parity scalar field. However, this equilibrium is inherently unstable \cite{Sanchis-Gual:2021edp}, though self-interactions similar to the Coleman Q-ball scalar potential have been found to stabilize the dipole \cite{Ildefonso:2023qty}. Beyond the dipole, chains of solitons have also been studied \cite{Sun:2022duv,Herdeiro:2020kvf}, including chains of rotating stars \cite{Gervalle:2022fze,Sun:2023ord} and self-interacting configurations \cite{Herdeiro:2021mol}. Similar multipolar configurations have also been shown to form dynamically when initial components are prepared with opposite charges that swap over time \cite{Jaramillo:2024smx, Copeland:2014qra, Xie:2021glp, Hou:2022jcd, Xie:2023psz}. Additionally, dipolar boson stars have a close relation with recent constructions of equilibrium systems involving two black holes \cite{Herdeiro:2023mpt, Herdeiro:2023roz}.

Notably, chains of boson stars, including dipolar configurations, do not have a flat space limit. The Klein-Gordon field alone cannot support dipolar static equilibrium solutions, even with generic self-interactions, as the gravitational attraction needed for balance is absent (see Sec. 5.2 of \cite{Cunha:2022tvk} and \cite{Shnir:2021gki} for a detailed argument). In contrast, other theories, such as the Yang-Mills monopole-antimonopole \cite{Kleihaus:1999sx}, Skyrmions \cite{Krusch:2004uf}, and others (reviewed in \cite{Shnir:2021gki}), do admit such solutions. Of particular interest are the even chains of spinning and charged Q-balls in the Maxwell-FLS model \cite{Loiko:2020htk}.

The purpose of this work is to demonstrate the existence of Einstein-FLS dipolar solutions, explore their stability within a region of parameter space using full 3D nonlinear evolutions, and show that flat space non-rotating Q-chains can also be constructed in the standard FLS model. The paper is organized as follows: in the next section, we present the Einstein-FLS model, its equations, and the ansatz. In Sec.~\ref{sec:static}, we construct the solutions and investigate their properties. The stability analysis of the dipolar solutions is addressed in Sec.~\ref{sec:dynamic}, and in Sec.~\ref{sec:Q-chains}, we construct static Q-dipoles, Q-chains and gravitating chains. We conclude with a discussion in Sec.~\ref{sec:conclusions}.

\section{Framework}\label{sec:framework}

We consider the 3+1 dimensional FLS model minimally coupled to Einstein's gravity, which describes a self-gravitating system consisting of a complex scalar field $\Phi$ and a real scalar field $\Psi$. The corresponding action is given by
\begin{equation}\label{eq:accion}
S=\int d^4 x\sqrt{-g} \left[\frac{1}{16 \pi G }R-g^{\mu\nu}\nabla_\mu\Phi\nabla_\nu\Phi^*-\frac{1}{2}g^{\mu\nu}\nabla_\mu\Psi\nabla_\nu\Psi-m^2\Psi^2|\Phi|^2-U(\Psi)\right],
\end{equation}
where $R$ is the Ricci scalar, $g$ the determinant of the spacetime metric $g_{\mu\nu}$ and $m$ is the coupling constant between the two scalar fields. The action contains the potential of the real scalar field 
\begin{equation}
  U(\Psi) = \mu^2\left(\Psi^2-v^2\right)^2 \, ,
\end{equation}
which introduces the positive parameter $\mu^2$ and $v$, the vacuum expectation value of $\Psi$. Throughout this paper we will use the signature $(-,+,+,+)$ and units where $c=1$.

According to Eq.~\eqref{eq:accion}, the complex scalar field $\Phi$ acquires mass through its coupling with the real scalar field $\Psi$, with the mass of $\Phi$ given by $mv$. In contrast, the mass of the real scalar field is $\sqrt{8}\mu v$.

The model described by Eq.~\eqref{eq:accion} is invariant under the global U(1) transformation $\Phi\to\Phi^{i\alpha}$. The associated conserved Noether current is $j_\mu = i\left(\Phi^*\nabla_\mu\Phi - \Phi\nabla_\mu\Phi^*\right)$ and the conserved charge is
\begin{equation}\label{eq:Q}
  Q = -\int j^0 \, \sqrt{-g}d^3 x \, .
\end{equation}

The field equations are obtained by varying $S$ with respect to the metric, the complex scalar field and the real scalar field.
Correspondingly we obtain the Einstein equations
\begin{eqnarray}\label{eq:Einstein}
&& R_{\mu\nu}-\frac{1}{2}R g_{\mu\nu}=8\pi G\  T_{\mu\nu};\\
&& T_{\mu\nu}=2\nabla_{(\mu}\Phi \nabla_{\nu)}\Phi^*+\nabla_{\mu}\Psi \nabla_{\nu}\Psi \nonumber \\ 
&& \qquad-g_{\mu\nu}\left[g^{\alpha\beta}\nabla_{\alpha}\Phi\nabla_{\beta}\Phi^*+\frac{1}{2}g^{\alpha\beta}\nabla_{\alpha}\Psi\nabla_{\beta}\Psi+m^2\Psi^2|\Phi|^2+\mu^2\left(\Psi^2-v^2\right)^2\right] \, ,\label{eq:Tmunu}
\end{eqnarray}
and the equations for the scalar fields
\begin{eqnarray}\label{eq:phi}
&&g^{\mu\nu}\nabla_\nu\nabla_\mu \Phi=m^2\Psi^2\ \Phi\,, \\
\label{eq:psi}
&&g^{\mu\nu}\nabla_\nu\nabla_\mu \Psi=2\left[m^2|\Phi|^2+2\mu^2\left(\Psi^2-v^2\right)\right]\Psi \, .
\end{eqnarray}

Interestingly, solitons in this model exist and can be constructed even in the limit where the potential vanishes, $\mu \to 0$, both in flat spacetime \cite{Loiko:2018mhb} and in curved spacetimes \cite{Kunz:2019sgn}. In these cases, $\Psi$ becomes massless and exhibits long-range behavior with a Coulomb-like asymptotic decay. In the opposite limit, where the real scalar field becomes infinitely massive, $\mu \to \infty$ with fixed $v$, $\Psi$ simplifies to $\Psi = v$, and the system described by Eq.~\eqref{eq:accion} reduces to the Einstein-Klein-Gordon (EKG) theory. This is a crucial observation, as we will begin our search for dipolar configurations in the full theory \eqref{eq:accion} by starting with the general relativistic dipolar solutions \cite{Yoshida:1997nd}.

The model in Eq.~\eqref{eq:accion} contains the parameters $m$, $\mu$, and $v$. However, the parameter $m$ can be eliminated (and thought of as setting the scale of the physical system), and the parameter $v$ can be reinterpreted as an effective gravitational coupling constant. To achieve this, we introduce the following dimensionless quantities \cite{Kunz:2019sgn}
\begin{equation}\label{eq:rescaling}
  \tilde{x}^\mu = mv x^\mu, ~~ \tilde{\Phi} = \frac{\Phi}{v} , ~~ \tilde{\Psi} = \frac{\Psi}{v}, ~~ \tilde{\mu} = \frac{\mu}{m} \, .
\end{equation}
Then, the equations of motion for the rescaled quantities are formally equivalent to the Eqs.~\eqref{eq:Einstein}-\eqref{eq:psi} after making the substitution $m=1$, $v=1$:
\begin{eqnarray}
  \label{eq:Einstein_R}
  && R_{\mu\nu}-\frac{1}{2}R g_{\mu\nu}=2\alpha^2\  T_{\mu\nu};\\
  \label{eq:phi_R}
  &&g^{\mu\nu}\nabla_\nu\nabla_\mu \Phi=\Psi^2\ \Phi\,, \\
  \label{eq:psi_R}
  &&g^{\mu\nu}\nabla_\nu\nabla_\mu \Psi=2\left[|\Phi|^2+2\mu^2\left(\Psi^2-1\right)\right]\Psi \, .
\end{eqnarray}
Here, we have dropped the tildes for simplicity. However, there is a key difference: the gravitational constant $G$ has been replaced by $\alpha^2/(4\pi)$, where $\alpha$ is a free parameter related to the vacuum expectation value through
\begin{equation}\label{eq:alpha}
    \alpha^2 = 4\pi G v^2.
\end{equation}
For the remainder of this paper, we will continue to omit the tildes for clarity. Consequently, the only two free parameters of the theory are $\mu$ and $\alpha$. In the next section, we will explore static solutions for various values of these parameters.

The gravitational decoupling limit is effectively achieved by considering Eqs.~\eqref{eq:Einstein_R}-\eqref{eq:psi_R} with $\alpha = 0$. However, it is important to note that in this flat spacetime limit, the physical quantities cannot be recovered using Eq.~\eqref{eq:rescaling}, as $\alpha = 0$ implies $v = 0.$ Despite this, the physical quantities can still be restored by choosing any value of $v$, since in the FLS model (in Minkowski space), $v$ can be rescaled out.

\section{Dipolar equilibrium solutions}\label{sec:static}

In the limit where the real scalar field $\Psi$ becomes infinitely massive, the system reduces to the EKG model, in which equilibrium solutions for two boson stars are known to exist \cite{Yoshida:1997nd}. These horizonless dipolar configurations are non-spinning and regular everywhere \cite{Cunha:2022tvk}. In line with these considerations, we adopt the following static and axisymmetric line element,
\begin{equation}\label{eq:metrica}
g_{\mu\nu}dx^\mu dx^\nu=-e^{2 F_0(r\theta)}dt^2+e^{2F_1(r,\theta)}(dr^2+r^2d\theta^2)+e^{2F_2(r,\theta)}r^2\sin^2\theta d\varphi^2,
\end{equation}
and for the scalar fields, we consider the following ansatz consistent with the previous assumptions
\begin{equation}
\Phi=\phi(r,\theta)e^{-i\omega t}\,,\quad \Psi=\psi(r,\theta) \, .
\end{equation}
The parameter $\omega$ is an unknown quantity in the model, determined by the boundary conditions. The five functions $F_i$ $(i=0,1,2)$, $\phi$, and $\psi$ depend on the radial coordinate $r$ and the polar angle $\theta$. The boundary conditions used to construct dipolar stars are specified as follows. Asymptotic flatness requires:
\begin{equation}\label{eq:out_bc}
  \begin{split}
    &\phi|_{r\to\infty}=0\,, \quad \psi|_{r\to\infty}=1 \, ;\\
    &F_0|_{r\to\infty}=0 \, , \quad F_1|_{r\to\infty}=0 \, , \quad F_2|_{r\to\infty}=0 \, ,
  \end{split}
\end{equation}
together with the condition $\omega<1$ in order to achieve the vacuum value for the complex field $\phi|_{r\to\infty}=0$.
Regularity of the solution at the origin and on the symmetry axis require,
\begin{equation}\label{eq:regularity_r}
  \begin{split}
    &\phi|_{r=0}=0\, , \quad \partial_r \psi|_{r=0}=0 \, ;\\
    &\partial_r F_0 |_{r=0}=0\, , \quad \partial_r F_1 |_{r=0}=0\, , \quad \partial_r F_2 |_{r=0}=0 \, ,
  \end{split}
\end{equation}
\begin{equation}\label{eq:regularity_th}
  \begin{split}
    &\partial_\theta \phi|_{\theta=0,\pi}=0 \,, \quad \partial_\theta \psi|_{\theta=0,\pi}=0\, ;\\
    &\partial_\theta F_0 |_{\theta=0,\pi}=0\,, \quad \partial_\theta F_1 |_{\theta=0,\pi}=0\,, \quad \partial_\theta F_2 |_{\theta=0,\pi}=0\,,\\
    &F_1|_{\theta=0,\pi}=F_2 |_{\theta=0,\pi} \, .
  \end{split}
\end{equation}
We impose that the spacetime remains invariant under reflection across the $\theta = \pi/2$ plane, while we impose the complex scalar field $\Phi$ to change sign. In this sense the complex scalar field is assumed to have odd parity with respect to a reflection along the equatorial plane $\theta = \pi/2$. Specifically, this condition implies that the derivatives of the metric functions with respect to $\theta$ vanish at $\theta = \pi/2$:
\begin{equation}\label{eq:reflection}
  \begin{split}
    & \phi|_{\theta=\pi/2}=0 \,, \quad \partial_\theta \psi|_{\theta=\pi/2}=0 \,;\\
    &\partial_\theta F_0 |_{\theta=\pi/2}=0 \, , \quad \partial_\theta F_1 |_{\theta=\pi/2}=0 \, , \quad \partial_\theta F_2 |_{\theta=\pi/2}=0 \, .
  \end{split}
\end{equation}
According to the previous $z=0$ plane symmetry conditions, the integration domain can be reduced to the region $0\leq\theta\leq\pi/2$.

Next, we present the explicit equations to be solved for the metric functions and the two scalar fields. These equations can be expressed in a compact form using the operators and combinations of the Einstein equations’ tensorial components as outlined in \cite{Bonazzola:1993zz}. To proceed, we first consider the 3+1 gravitational source terms $\rho = T_{\mu\nu} n^\mu n^\nu$ (energy density) and $S_{\alpha \beta} = T_{\mu\nu} \gamma^\mu_\alpha \gamma^\nu_\beta$ (stress tensor), with $n^\alpha = (e^{-F_0}, 0, 0, 0)$ and $\gamma^\alpha_\beta = \delta^\alpha_\beta + n^\alpha n_\beta$; both geometrical quantities properly defined in Sec~\ref{sec:dynamic}. The Einstein equations for the metric \eqref{eq:metrica} are then given by:
\begin{eqnarray}
\Delta_3 F_0 &= &4\pi G A^2 \left(\rho+S\right)  - \partial F_0 \partial\left( F_0 + F_2\right) \label{eq:nu} \\
\Delta_2 \left[\left(NB -1\right)r \sin\theta\right] & = &  8\pi G N A^2 B r \sin\theta \left(S^r_{\ \, r} + 
    S^\theta_{\ \, \theta} \right) 
    \label{eq:B} \\
\Delta_2\left(F_1  +  F_0\right) & = & 8\pi G A^2 S^\varphi_{\ \, \varphi} - \partial  F_0 \partial  F_0 ,  \label{eq:A}
\end{eqnarray}
where $N=e^{F_0}$ (lapse function), $A=e^{F_1}$, $B=e^{F_2}$ and the operators $\Delta_i$ correspond to the i-dimensional flat Laplacian and $\partial f\partial g$ is the product of the gradients of $f$ and $g$. All of these operators acting on two dimensional $r,\theta$ dependent functions.
%

To determine the explicit form of the source terms based on the ansatz for the metric and fields, we derive the following expressions:
\begin{eqnarray}
\rho + S &=& 4\frac{\omega^2}{N^2} \phi^2 - 2\psi^2\phi^2 - 2\mu^2\left(\psi^2-1\right)^2 \, ,\label{e:EpS}\\
S^r_{\ \, r} + S^\theta_{\ \, \theta} & = &  2 
    \frac{\omega^2}{N^2}
    \phi^2  - 2\psi^2\phi^2 - 2\mu^2\left(\psi^2-1\right)^2 \, , \label{e:Srr_Sthth}\\
S^\varphi_{\ \, \varphi} & = & 
     \frac{\omega^2}{N^2}
    \phi^2 - \frac{1}{A^2} \partial \phi \partial \phi - \frac{1}{2 A^2} \partial \psi \partial \psi - \psi^2\phi^2-\mu^2\left(\psi^2-v^2\right)^2 . 
\end{eqnarray}

Using the differential operators previously introduced, the equation of motion for $\Phi$ and $\Psi$ reduce to
\begin{equation}\label{eq:phi_2D}
\Delta_3\phi=A^2\left(\psi^2-\frac{\omega^2}{N^2}\right)\phi-\partial\phi\partial(F_0+F_2)
\end{equation}
and 
\begin{equation}\label{eq:psi_2D}
\Delta_3\psi=2A^2\left[\phi^2+2\mu^2\left(\psi^2-v^2\right)\right]\psi-\partial\psi\partial(F_0+F_2) \, .
\end{equation}

The resulting set of five coupled nonlinear elliptic differential equations, \eqref{eq:nu}-\eqref{eq:B}, \eqref{eq:phi_2D}, and \eqref{eq:psi_2D}, subject to the boundary conditions \eqref{eq:regularity_r}, \eqref{eq:regularity_th}, and \eqref{eq:reflection}, is solved numerically. We employ the spectral methods library \texttt{Kadath}~\cite{Kadath,Grandclement:2009ju} to solve this elliptic system. In the solver, we utilize the \texttt{Polar} space module, which facilitates the compactification of $r$. For all dipolar configurations in this work, we used 17 Chebyshev spectral coefficients in both the $r$ and $\theta$ directions, distributed across 8 radial domains with boundaries at $r=2^i$ $(i=0,1,\cdots,6)$. Beyond dipolar cases (as those presented in Sec.~\ref{sec:Q-chains}) we have employed the same radial domains but 25 spectral coefficients in some cases.

The solver employs an iterative Newton-Raphson scheme, which requires an initial guess for the functions to initiate the iterative process and ensure convergence to a solution. This process is delicate, however we found that the following initial guess enables a successful iterative process when $\mu \to \infty$, allowing the system to converge to the EKG dipolar boson star solution:
\begin{equation}
\phi=r \phi_0  Y_{1}^{0}(\theta,\varphi) e^{-x^2/\sigma_x^2-z^2/\sigma_z^2} \, , \quad N=1-(1-N_0)e^{-r^2}, \quad \omega=0.965,
\end{equation}
where $\phi_0 = 10^{-2}$, $x=r\sin\theta$, $z=r\cos\theta$, $\sigma_x^2 = 200$, $\sigma_z^2 = 50$, and $Y_{1}^{0}(\theta,\varphi)$ is the spherical harmonic with $\ell = 1$ and $m = 0$. In this process, $\omega$ is treated as an unknown, and to prevent convergence to the trivial vacuum solution, we impose that the lapse function $N$ at the center of the dipole satisfies $N(r=0) = N_0 = 0.9404$. The iterative process is terminated and a solution is obtained when the residual drops below $10^{-8}$.

Once the first solution is obtained, the remaining dipolar boson star solutions in the theory \eqref{eq:accion} can be generated by gradually varying the theory’s parameters $(\alpha, \mu)$ and the value of the lapse function at the origin. We found that the lapse function at the center is a useful quantity for parametrizing sequences of solutions with fixed $(\alpha, \mu)$. In this way, we constructed the equilibrium family of two-boson star solutions in the EKG theory.

For finite $\mu$, solutions are obtained by using a corresponding equal-frequency EKG solution as an initial guess and gradually decreasing the values of $\mu$ and $\alpha$. Each solution has a specific value of charge, determined by Eq.~\eqref{eq:Q}. The other key quantities that characterize the dipolar stars are the total mass and the separation between the two components. We extract the value of the mass by comparing the numerical solutions with the asymptotic sub-leading expansion for the lapse function,
\begin{equation}\label{eq:mass_from_lapse}
  N^2 = e^{2F_0} = 1 - \frac{\alpha^2 M}{2\pi r} + \mathcal{O}\left(\frac{1}{r^2}\right) \, .
\end{equation}
To facilitate comparison with the previously reported $\mu \to \infty$ solutions \cite{Herdeiro:2020kvf, Sanchis-Gual:2021edp, Cunha:2022tvk}, we will follow the approach presented in \cite{Herdeiro:2023lze} and report the quantity $M\,Gv$.\footnote{
Alternatively, one could achieve this by setting $\alpha^2 = 0.25$ and noting that in the $\mu \to \infty$ limit, the constant $\alpha$ can be rescaled away. By rescaling to solutions with $\alpha^2 = 4 \pi$, i.e., in units where $G=1$ according to Eq.~\eqref{eq:alpha}, we have verified that the total mass for the sequence of solutions reported in \cite{Herdeiro:2020kvf, Sanchis-Gual:2021edp, Cunha:2022tvk} for the dipole boson star is reproduced.
}
In addition to the expression \eqref{eq:mass_from_lapse}, we have used the Komar integral \cite{Wald:1984rg,Gourgoulhon:2010ju}
\begin{equation}\label{eq:komarM}
  M = - \int \left( 2{T_{t}}^t  - T \right)\,\sqrt{-g} d^3 x \,,
\end{equation}
to check for numerical accuracy of the solutions. The relative errors between the two definitions of mass are on the order of $10^{-6}$. Solutions were discarded whenever this error indicator exceeded $10^{-3}$.

\begin{figure}
  \includegraphics[width=0.6\textwidth]{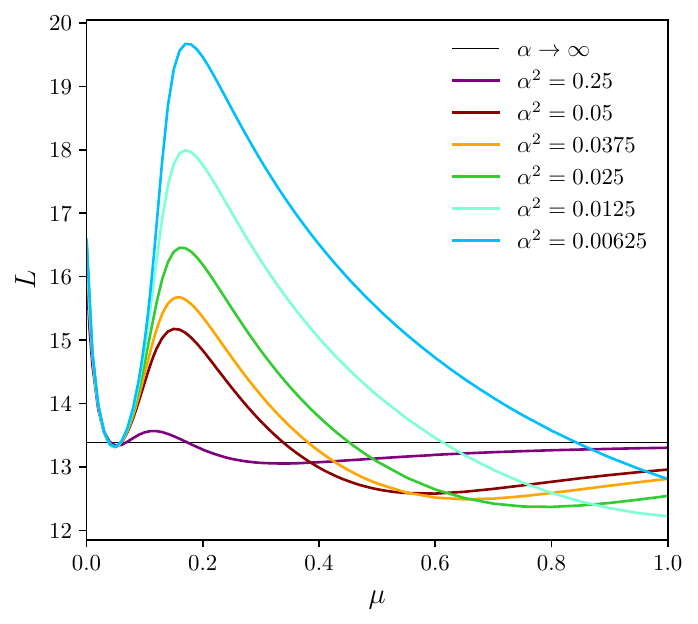}
    \caption{
      Proper distance $L$ between the dipolar components as a function of the real scalar field mass $\mu$ for a fixed frequency of $\omega = 0.9$ and various values of $\alpha$. As $\mu$ approaches infinity, all curves converge to the separation corresponding to the EKG dipole, as expected. Similarly, for fixed $\mu$, the limit as $\alpha$ approaches infinity also tends toward the EKG star dipole. A thin line has been added to indicate this limit.
    }
    \label{fig:L}
\end{figure}

The separation between the dipolar components can be determined by locating the symmetric maxima of the energy density $\rho$, which are situated along the $z$-axis at positions with $r = z_0$. Thus, the relevant quantity related to the separation of the two stars is defined as the proper distance between the points $z = \pm z_0$:
\begin{equation}
  L = 2\int_0^{z_0} e^{F_1(r,0)} dr \, .
\end{equation}
We begin our analysis by examining the influence of the model parameters on the properties of the solutions. To this end, we fix the frequency of the scalar field oscillations to $\omega = 0.9$; the significance of this choice will become clear shortly. In Fig.~\ref{fig:L}, we present the separation as a function of the real scalar field mass $\mu$ for various values of the gravitational coupling $\alpha$.

The first observation from Fig.~\ref{fig:L} is that solutions exist even in the case of a vanishing potential, similar to the spherically symmetric case reported in \cite{Loiko:2018mhb}. Additionally, the configurations appear to converge to the same value of $L$ in the $\mu \ll 1$ region. We have examined other quantities in this limit and found that they rescale according to specific relations, such as $M = c_1\mu + c_2$ and $Q = c_3\mu + c_4$ with $c_1$, $c_2$, $c_3$ and $c_4$ constants that depend slightly in the specific value of $\alpha$. The constants $c_3$ and $c_4$ are within the interval $(15.5,16.2)$. Beyond this region, the value of $L$ increases, reaching a local maximum before decreasing toward the finite EKG limit, which coincides with the limit $\alpha \to \infty$. Notably, the maximum separation near $\mu = 0.2$ seems to increase without bound as $\alpha \to 0$. This observation will be important in Sec.~\ref{sec:Q-chains}, where we discuss the existence of chain solutions in flat spacetime.

\begin{figure}
\includegraphics[width=0.50\textwidth]{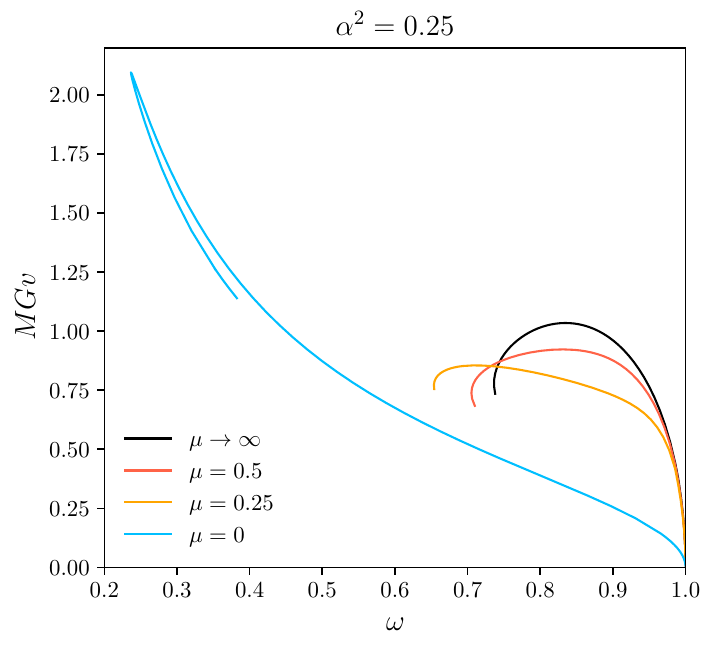} \includegraphics[width=0.49\textwidth]{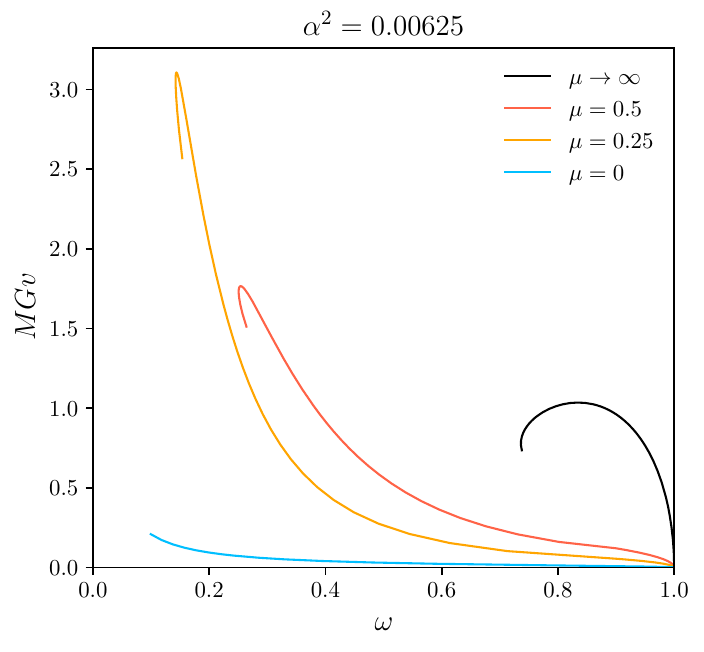}
  \caption{
  Mass-frequency diagram for FLS dipoles for three selected cases of $\mu$, along with the Klein-Gordon limit as $\mu \to \infty$. (Left panel) The sequence with a strong gravitational coupling, $\alpha^2 = 0.25$, which corresponds to the highest $\alpha$ value shown in Fig.~\ref{fig:L}. (Right panel) The sequence with a weak gravitational coupling, $\alpha^2 = 0.00625$, corresponding to the lowest $\alpha$ value considered in Fig.~\ref{fig:L}.
  }
  \label{fig:M}
\end{figure}

Many, bosonic star configurations that possess stable solutions against small perturbations lie along distinctive branches, clearly separated from unstable branches in their domain of existence (see e.g., the reviews \cite{Liebling:2012fv,Bezares:2024btu} on the dynamics of boson stars). These branches are often depicted in mass-frequency diagrams. In such diagrams, particularly for polynomial complex scalar field potentials, the frequency $\omega$ has a global minimum, and gravitationally stable configurations are located in regions where $dM/d\omega < 0$ and between the Newtonian limit and the global minimum of $\omega$ (see \cite{Gleiser:1988ih, Jetzer:1989us, Sanchis-Gual:2017bhw, Siemonsen:2020hcg} for some examples). This pattern also holds true for self-interacting dipolar stars \cite{Ildefonso:2023qty}. Thus, it is instructive to examine the mass-frequency diagrams of the solutions presented here. In Fig.~\ref{fig:M}, we display the mass-frequency diagrams for two cases of coupling to gravity $\alpha$ and the selected values of the real scalar field mass $\mu = 0$, 0.25, and 0.5. In particular we present the full sequence of solutions for both the largest and smallest values of $\alpha$ considered in Fig.~\ref{fig:L}. With reasonable confidence, we establish that all configurations plotted for $L$, with $\omega = 0.9$, reside along the \textit{candidate} stable branch, connecting the $(M = 0, \omega = 1)$ point with the critical mass configuration in each case.

Also from comparing both panels in Fig.~\ref{fig:M}, we confirm that for a given value of $\mu$, the solutions deviate more from the EKG limit as $\alpha$ decreases. Additionally, the range of frequency values $\omega$ for which solutions exist expands as both $\mu$ and $\alpha$ decrease. In particular, for the case $\mu = 0$, $\alpha^2 = 0.00625$ (right panel, blue line), we could not identify the critical mass configuration or the minimum frequency solution. Beyond the minimum $\omega$ obtained and presented in the plot for this sequence, the error indicator obtained by comparing the mass definitions \eqref{eq:mass_from_lapse} and \eqref{eq:komarM} exceeded the 0.1\% of relative difference, and after a few more points with smaller $\omega$, the code could no longer converge.

\begin{figure}
\includegraphics[width=0.49\textwidth]{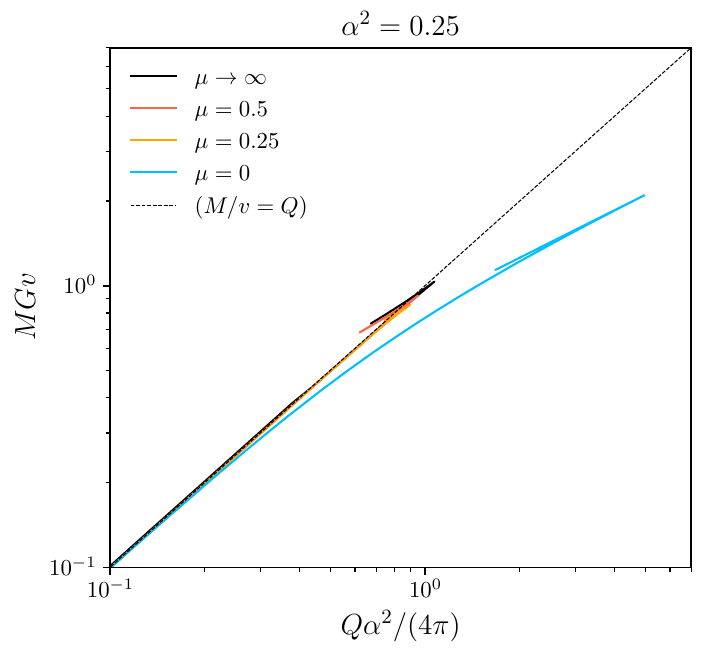} \includegraphics[width=0.49\textwidth]{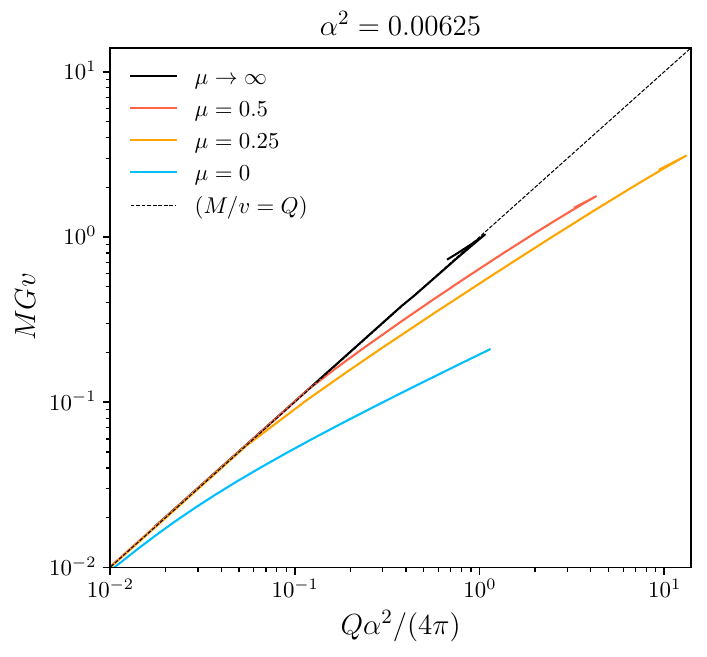}
\caption{
  Energy $M$ of the dipolar FLS sequences as a function of the charge $Q$, corresponding to the same cases as in Fig.~\ref{fig:M}. Configurations below the thin dashed line, where $M/v = Q$, are gravitationally bound. In all explored cases, the configurations within the region connecting the $M = 0$ solution to the critical mass configuration—i.e., those in the candidate stable branch—are gravitationally bound.
  }
  \label{fig:Q}
\end{figure}

Some insight into the potential stability of the solutions can also be gained from mass-charge ($M$-$Q$) diagrams. Classical stability criteria for Q-balls have been applied to boson stars \cite{Liebling:2012fv} to establish a necessary condition for equilibrium configurations to be gravitationally bound and unlikely to fragment under small perturbations. It is important to note, however, that gravitationally unbound yet stable configurations have been found for a logarithmic scalar potential \cite{Jaramillo:2024smx}. Additionally, there are numerous examples of gravitating systems that, while classically stable in the sense that $m_b Q > M$, with $m_b$ the boson mass, are actually unstable. For instance, dipolar boson stars in the EKG theory exhibit such behavior \cite{Sanchis-Gual:2021edp, Ildefonso:2023qty}. In Fig.~\ref{fig:Q}, the $M(Q)$ curves for the dipolar FLS model are displayed, corresponding to the same configurations as in Fig.~\ref{fig:M}.
We see that all solutions in the candidate stable branch, in the line connecting $M=0$ with $\max(M)$ in each case, satisfy the condition $M < v Q$. Therefore correspond to configurations with negative binding energy and are not expected to fission under the effect of small perturbations. We will see in the next section that unlike EKG dipolar boson stars, configurations in the candidate stable branch are not always unstable.

After mapping out the domain of existence for dipolar stars in the Einstein-FLS theory, we now explore the stability of these configurations by evolving a broad range of cases. While we confirm that the two previously established stability criteria are necessary for achieving perturbatively stable solutions, we find that they are not sufficient for all values of the parameters $\mu$ and $\alpha$.

\section{Dynamical evolution and stability}\label{sec:dynamic}

In this section, we aim to explore the dynamical stability of the previously constructed isolated static solutions. By evolving these configurations, we can assess their response to perturbations and understand their long-term behavior. To achieve this, we will use the static solutions as initial data for our dynamical simulations. These systems will be subjected to small perturbations arising from the truncation errors inherent in the numerical implementation, and additional perturbations will be introduced by hand to study non-static configurations from the onset. This approach allows us to thoroughly test the stability of the dipolar stars through the model parameter space.

\subsection{Setup}\label{sec:setup}

To track the system’s dynamics, we implemented a numerical code capable of solving the Einstein-FLS system. This was accomplished using the \textsc{Einstein Toolkit} framework~\cite{EinsteinToolkit:web,Loffler:2011ay, Zilhao:2013hia}, with mesh refinement handled by the \textsc{Carpet} package~\cite{Schnetter:2003rb}.
As a preparatory step, we performed a Cauchy (3+1) decomposition of the scalar field equations of motion Eqs.~\eqref{eq:phi}, \eqref{eq:psi} and evolve initial data in the full two-component Einstein-FLS model. We cast the field equations into the 3+1 form by considering the spacetime line element
\begin{equation}
  g_{\mu\nu} dx^\mu dx^\nu = -(N^2-\beta^i\beta_i)dt^2 + 2\gamma_{ij}\beta^i dtdx^j + \gamma_{ij}dx^idx^j \,, 
\end{equation}
and introducing the extrinsic curvature, defined as
\begin{equation}
  K_{ij} = -\frac{1}{2N}(\partial_t - \mathcal{L}_\beta)\gamma_{ij} \, , 
\end{equation}
where $N$ is the lapse function, $\beta^i$ is the shift vector, $\gamma_{ij}$ is the metric defined on the hypersurfaces of constant coordinate $t$ and $\mathcal{L}$ is the Lie derivative.

The evolutions were performed by adapting an existing ``thorn'' in the toolkit, which solves for multiple complex scalar fields using finite differences~\cite{Sanchis-Gual:2019ljs,Jaramillo:2020rsv,Jaramillo:2022zwg}. The FLS system was numerically integrated using fourth-order spatial discretization provided by the toolkit's infrastructure. This toolkit solves the Einstein equations within the ADM 3+1 framework, utilizing the Baumgarte-Shapiro-Shibata-Nakamura (BSSN) formulation for evolving spacetime variables \cite{Nakamura:1987zz,Shibata:1995we}. The \textsc{McLachlan} thorn~\cite{Brown:2008sb} was used to evolve spacetime, with matter source terms derived from the energy-momentum tensor \eqref{eq:Tmunu} for the complex and real scalar field coupled system.

For all evolutions, we applied the 1+log slicing condition for the lapse function $N$ and the Gamma-driver condition for the shift vector $\beta^{i}$  \cite{alcubierre2008introduction}. To evolve the scalar fields $\Phi$ and $\Psi$ over time, $\Phi$ was decomposed into its real and imaginary components 
\begin{equation}
    \Phi = \Phi_r + i\Phi_i \, ,
\end{equation}
and the BSSN formulation was coupled to the evolution equations for these components and the real field, yielding
\begin{subequations}
    \begin{eqnarray}
\left(\partial_t - \mathcal{L}_{\beta} \right)& \Phi_r & = - 2 N K_{\Phi_r} \,, \\
  \left(\partial_t - \mathcal{L}_{\beta} \right)& \Phi_i & = - 2 N K_{\Phi_i} \,, \\
  \left(\partial_t - \mathcal{L}_{\beta} \right)& \Psi   & = - 2 N K_{\Psi}   \,, \\
    \end{eqnarray}
\end{subequations}
which define a the canonical momentum variables associated to the scalar fields and evolve according to the equations
\begin{subequations}\label{eq:bssn_fields}
\begin{eqnarray}
  \left(\partial_t - \mathcal{L}_{\beta} \right)& K_{\Phi_r} &  = N \left[ K K_{\Phi_r} - \frac{\chi}{2} \tilde{\gamma}^{ij} \tilde{D}_i \partial_j \Phi_r + \frac{1}{4} \tilde{\gamma}^{ij} \partial_i \Phi_r \partial_j\chi
                  + \frac{1}{2}  \Psi^2\Phi_r \right]\nonumber \\
                  && ~~- \frac{\chi}{2} \tilde{\gamma}^{ij} \partial_i N \partial_j \Phi_r \,, \\
  \left(\partial_t - \mathcal{L}_{\beta} \right)& K_{\Phi_i} &  = N \left[ K K_{\Phi_i} - \frac{\chi}{2} \tilde{\gamma}^{ij} \tilde{D}_i \partial_j \Phi_i + \frac{1}{4} \tilde{\gamma}^{ij} \partial_i \Phi_i \partial_j\chi
                 + \frac{1}{2} \Psi^2 \Phi_i \right]\nonumber \\
                 && ~~- \frac{\chi}{2} \tilde{\gamma}^{ij} \partial_i N \partial_j \Phi_i \,,  \\
  \left(\partial_t - \mathcal{L}_{\beta} \right)& K_{\Psi} &  = N \left[ K K_{\Psi} - \frac{\chi}{2} \tilde{\gamma}^{ij} \tilde{D}_i \partial_j \Psi + \frac{1}{4} \tilde{\gamma}^{ij} \partial_i \Psi \partial_j\chi
                + (\Phi_r^2+\Phi_i^2+2\mu^2(\Psi^2-1))  \Psi \right] \nonumber \\
               && ~~- \frac{\chi}{2} \tilde{\gamma}^{ij} \partial_i N \partial_j \Psi \,,  
\end{eqnarray}
\end{subequations}
To solve these equations we employed the \texttt{MoL} thorn, which utilizes a fourth-order Runge-Kutta scheme for time integration. In the Eqs.~\eqref{eq:bssn_fields}, $K$  represents the trace of the extrinsic curvature, and the BSSN formulation introduces the conformal factor $\chi$, defining a conformal spatial metric $\tilde{\gamma}_{ij}$ with unit determinant via the relation $\tilde{\gamma}_{ij} = \chi \gamma_{ij}$. The formulation also introduces $\tilde{D}$, the covariant derivative compatible with this conformal metric. These adjustments support (numerically) stable evolution and accurate coupling between matter and geometry in our simulations.

Mesh refinement was managed using \textsc{Carpet}. The fixed mesh refinement grid hierarchy consisted of nested rectangular domains with three levels of refinement, with the finest level encompassing the entire configuration. The physical grid spans a cube with a side length of 200 units. Since all fields involved are axisymmetric, we impose symmetry along both the $x$ and $y$ axes. Given that the matter clumps are aligned along the z-axis, we extend the inner refinement levels to be 1.5 times longer in the $z$-dimension than in the $x$ and $y$ dimensions. To ensure the star’s properties are accurately captured, we set the spatial resolution on the finest refinement level to $\Delta x^i = 0.5$, and on the coarsest level to $\Delta x^i = 4.0$.

\subsection{Evolutions}\label{sec:evolutions}

%
\begin{figure}
  \includegraphics[width=0.45\textwidth]{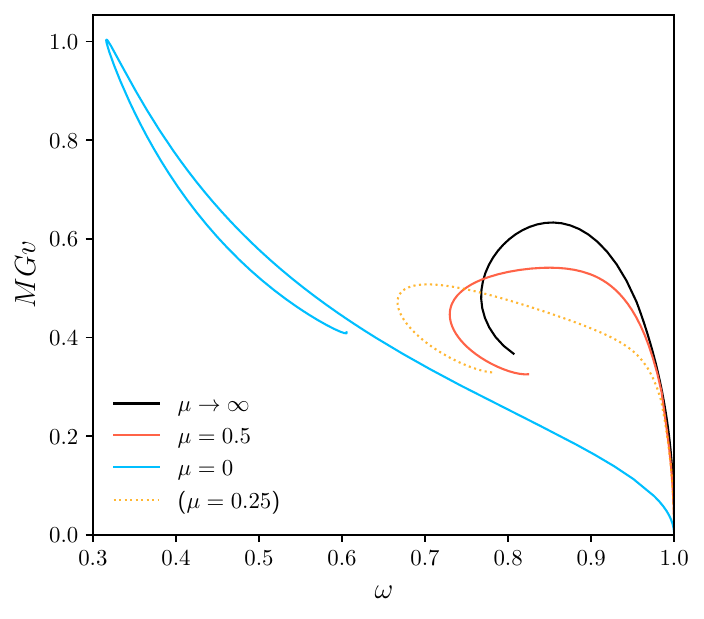} \includegraphics[width=0.45\textwidth]{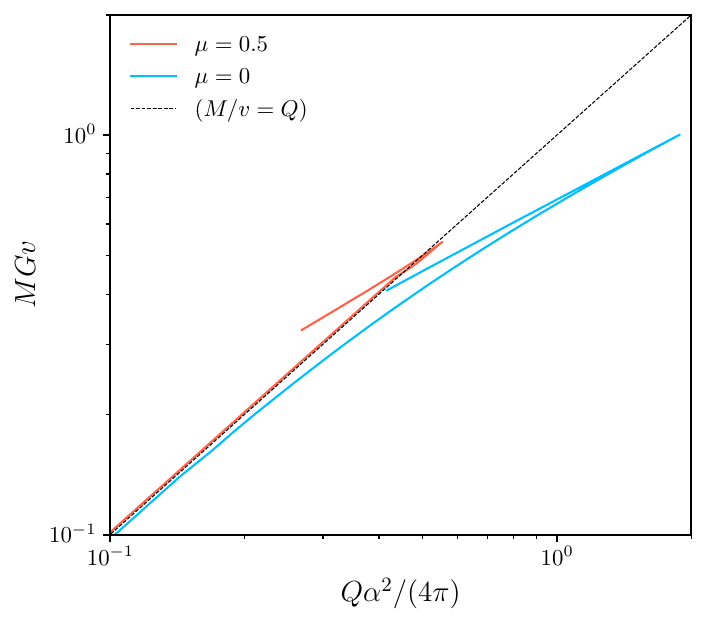}
    \caption{
    Spherical Klein-Gordon and FLS boson stars with $\alpha^2=0.25$.
    }
    \label{fig:MQ-spherical}
\end{figure}
\begin{table}[b]
  \centering
  \begin{tabular}{lcc|cc|c}
    \hline
$\alpha^2$ & $\omega$ & $\mu$ & $M$ & $Q$ & Result \\ 
\hline
0.25 & 0.9 & 0.5   & 0.527 & 0.539 & Stable  \\
0.25 & 0.9 & 0.25  & 0.418 & 0.426 & Stable  \\
0.25 & 0.9 & 0     & 0.166 & 0.172 & Stable  \\
0.25 & 0.8 & 0.25  & 0.476 & 0.493 & Stable  \\
0.25 & 0.8 & 0     & 0.254 & 0.275 & Stable  \\
\hline
\end{tabular}
\caption{Spherical FLS boson stars. All these configurations are on the candidate stable branch and are gravitationally bounded.}
\label{tab:spherical}
\end{table}
As a preliminary sanity check, before proceeding with the dipolar star evolutions, we first consider spherical boson stars, which are not expected to deviate from the established stability scenario for free scalar field spherical boson stars (in particular in the $\mu\gg1$ or $\alpha\gg1$ cases). It is worth noting, however, that no stability analysis has been conducted for these configurations in the literature. Nevertheless, the static spherical solutions were presented in \cite{Kunz:2019sgn} and further analyzed in \cite{deSa:2024dhj}, as mentioned in the introduction.

We have constructed spherical boson stars\footnote{
  The spherical configurations can be obtained from the equations presented in Sec.~\ref{sec:static} for parity-even $\Phi$, substituting the boundary condition $\phi|_{r=0}=0$ and $\phi|_{\theta=\pi/2}=0$ by $\partial_r \phi|_{r=0}=0,\quad \partial_\theta \phi|_{\theta=\pi/2}=0$. The two-dimensional solver converges to the spherical symmetric boson star in the FLS model. We reproduce results presented in Fig.~1 of \cite{Kunz:2019sgn}.
}.
and present the equilibrium sequence of them in Fig.~\ref{fig:MQ-spherical} for a single value of $\alpha$ and
$\mu$ between 0 and 0.5. As expected that configurations differ more from the mini-boson star monotonically with decreasing value of $\mu$. 
According to the right plot of Fig.~\ref{fig:MQ-spherical} all configurations between the zero mass solution and the minimum frequency solution are gravitationally bound. In particular this means all solutions $\omega = 0.9$ are not expected to disperse.
Next, we select several solutions, shown in Table~\ref{tab:spherical} which are located in the stable branch, and use the stationary solutions obtained from the spectral elliptic solver to generate initial data for $\Phi$, $\Psi$, and $g_{\mu\nu}$ (along with their corresponding canonical momenta). This is achieved by interpolating the spectral coefficients onto the Cartesian grid of the evolution code. We then evolve the system up to $t = 10^4$, allowing the truncation error from the numerical code to naturally perturb the star. None of these configurations exhibit any signs of instability, even those that differ significantly from the EKG case in both their scalar and gravitational components. Confident in the presented approach for constructing and evolving these configurations, we proceed to explore the dipolar configurations, which possess markedly different dynamical properties.

\begin{table}[b]
  \centering
  \begin{tabular}{lcc|ccc|c}
    \hline
$\alpha^2$ & $\omega$ & $\mu$ & $L$ & $M$ & $Q$ &  Result \\ 
\hline
-    & 0.9 &$\infty$ & 13.39 & 0.961 & 0.989 & Unstable \\
$4\pi$ & 0.9 & $\{0,0.25,\mathbf{0.5}\}$   & 13.38 & 0.960 & 0.987 & Unstable \\
0.25 & 0.9   & $\{0,\mathbf{0.25},0.5\}$   & 13.13 & 0.729 & 0.744 & Unstable \\
0.25 & 0.8   & $\{\mathbf{0},0.25\}$       & 12.05 & 0.389 & 0.422 & Unstable \\
0.1  & 0.9   & $\{0,\mathbf{0.25},0.5\}$   & 13.43 & 0.505 & 0.513 & Unstable \\
\hline
\end{tabular}
\caption{Dipolar FLS boson stars. Broad scan for the stability of twelve configurations located on their corresponding candidate stable branch, see Figs.~\ref{fig:M} and \ref{fig:Q}. The quantities $L$, $M$ and $Q$ correspond to the solution with value of $\mu$ in bold of the corresponding row.}
\label{tab:dipolar}
\end{table}
We confirm that dipolar boson stars near the EKG limit are unstable. In the following analysis, we focus primarily on the cases with $\omega = 0.9$, unless otherwise noted, in which case we consider $\omega = 0.8$ for generality. For instance, several of the dipolar configurations listed in Table~\ref{tab:dipolar} can be shown to be close, or very close, to their corresponding EKG counterparts. In particular, the three configurations with $\alpha^2 = 4\pi$ display global quantities nearly identical to those of the dipolar mini-boson star. We evolved these configurations for three values of the real scalar field mass, $\mu=0$, 0.25, and 0.5, finding all of them to be unstable.

Next, we conducted a broad scan of the parameter space, focusing on the cases where $\alpha^2=0.25$ and $\alpha^2=0.1$, which already deviate from the EKG limit. These cases also proved to be unstable. We further tested cases with $\omega = 0.8$, since this frequency lies within the candidate stable branch when $\mu$ decreases (as seen in Fig.~\ref{fig:M}, particularly for $\alpha^2=0.25$). However, these configurations also turned out to be unstable. All these configurations start to develop an instability before $t\sim2000$ but not after $t\sim500$.

\begin{figure}
  \subfigure[~~Stable configuration. Magnitude of the complex scalar field $|\Phi|^2$]{
  \includegraphics[height=0.13\textheight]{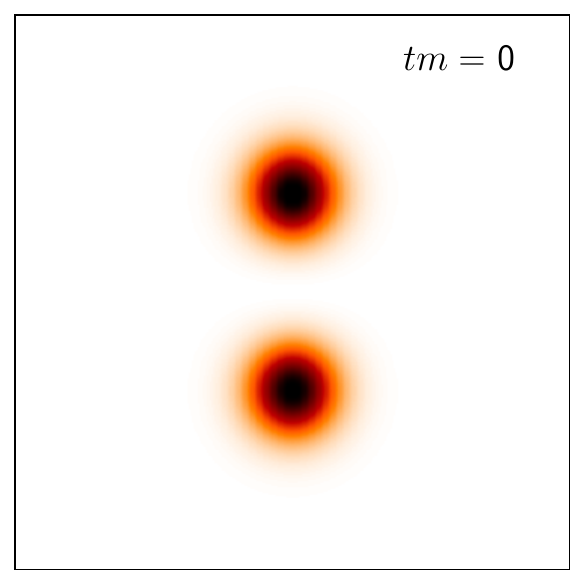}\hspace{-0.1cm}
  \includegraphics[height=0.13\textheight]{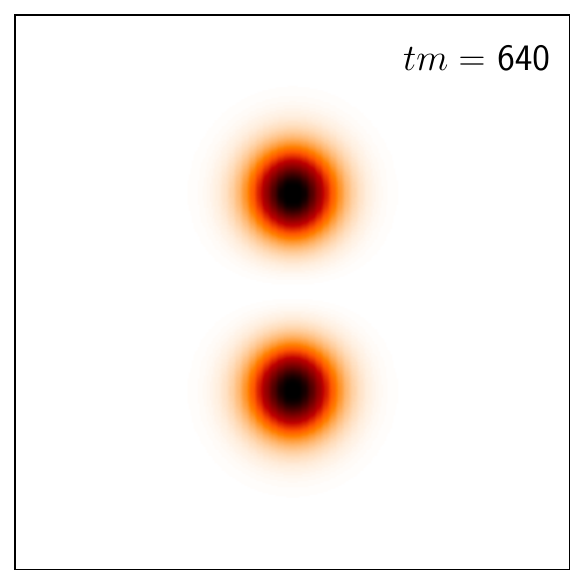}\hspace{-0.1cm}
  \includegraphics[height=0.13\textheight]{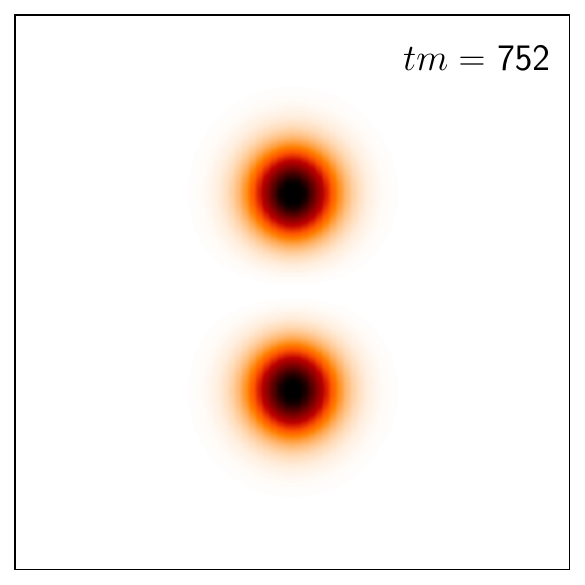}\hspace{-0.1cm}
  \includegraphics[height=0.13\textheight]{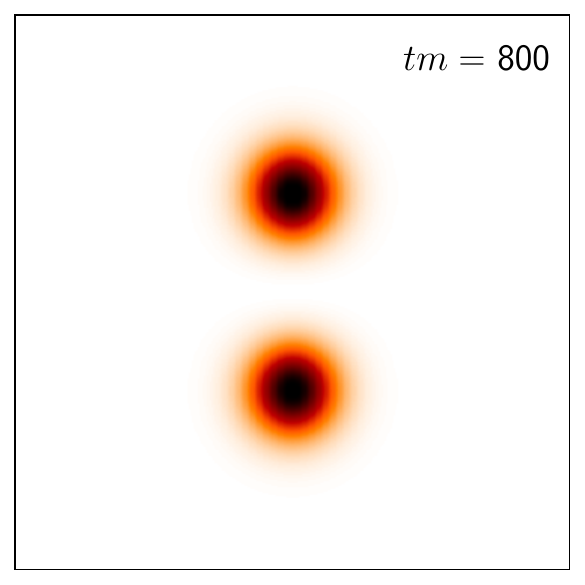}\hspace{-0.1cm}
  \includegraphics[height=0.13\textheight]{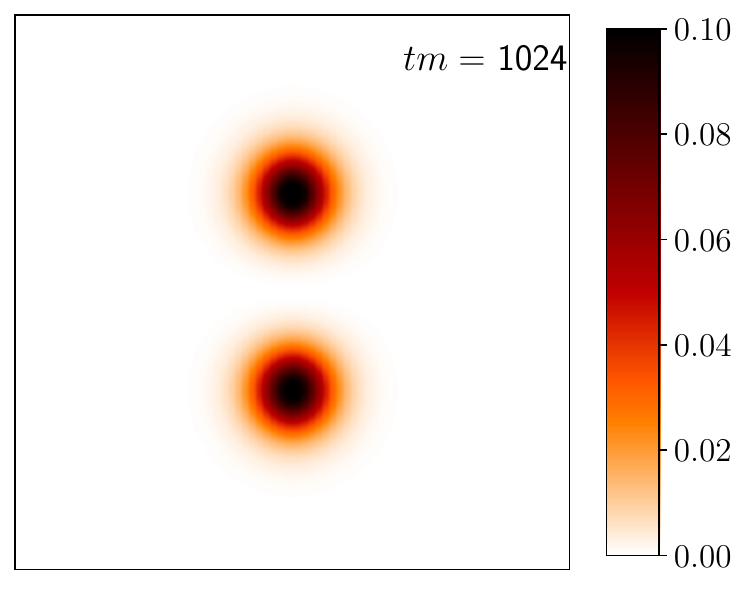}
  }
  \subfigure[~~Unstable configuration. Magnitude of the complex scalar field $|\Phi|^2$]{
  \includegraphics[height=0.13\textheight]{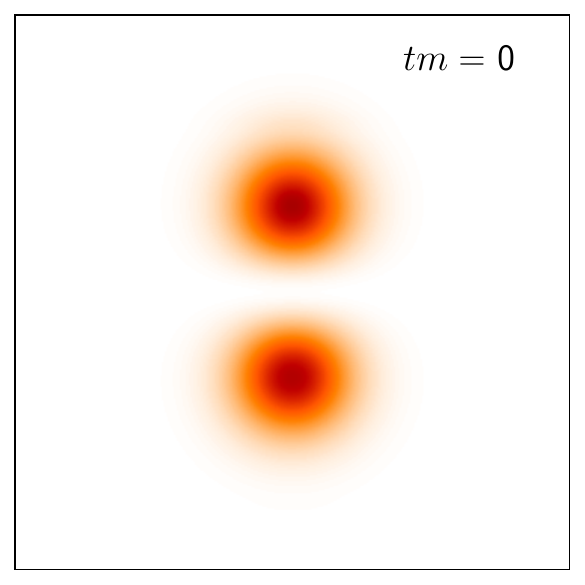}\hspace{-0.1cm}
  \includegraphics[height=0.13\textheight]{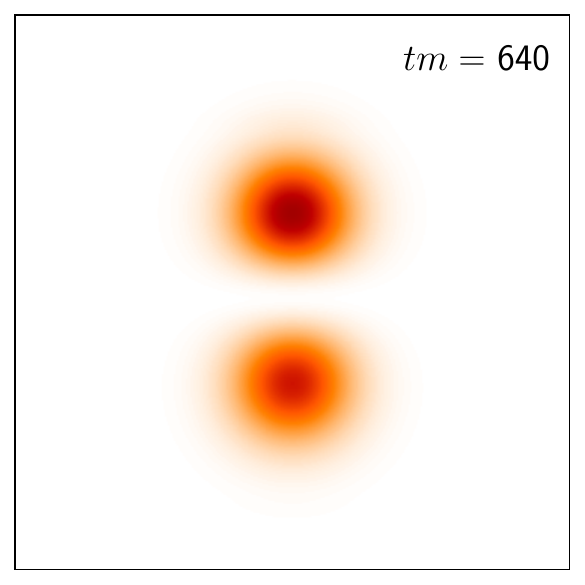}\hspace{-0.1cm}
  \includegraphics[height=0.13\textheight]{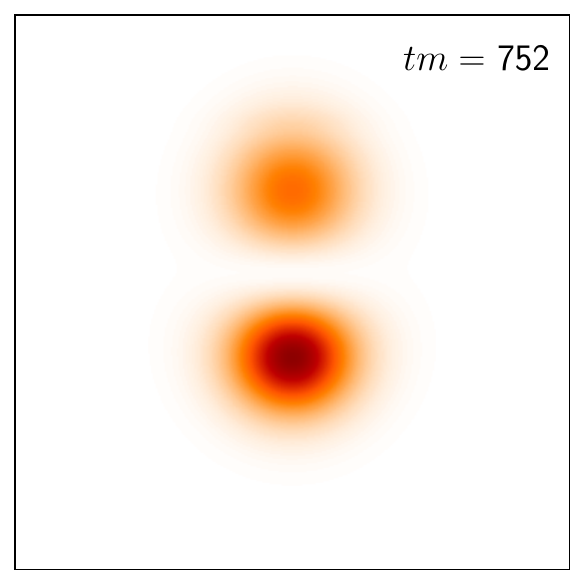}\hspace{-0.1cm}
  \includegraphics[height=0.13\textheight]{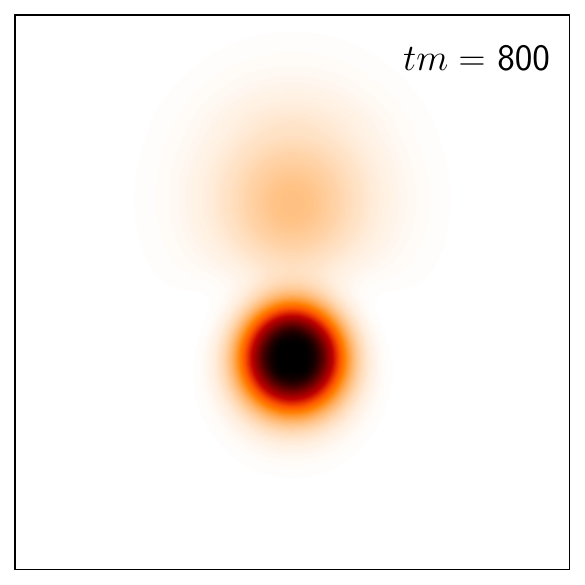}\hspace{-0.1cm}
  \includegraphics[height=0.13\textheight]{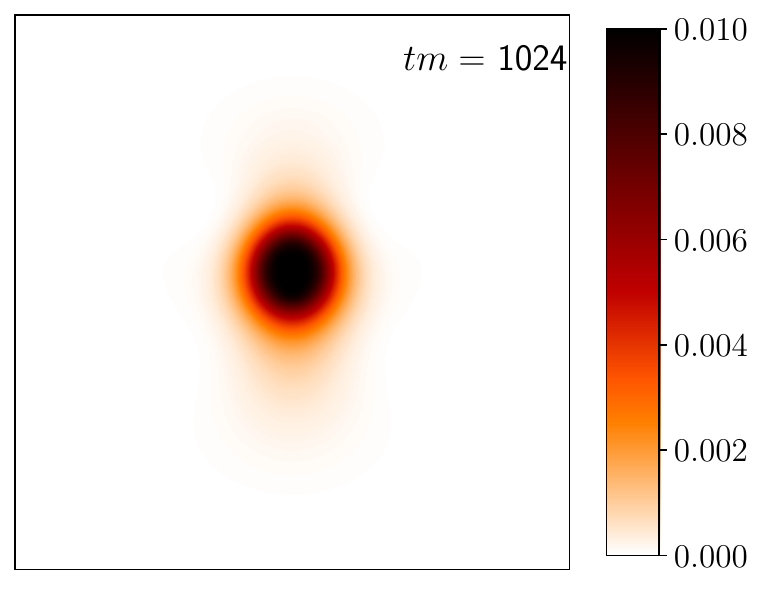}
  }
  \subfigure[~~Real scalar field $\Psi$]{
  \includegraphics[height=0.13\textheight]{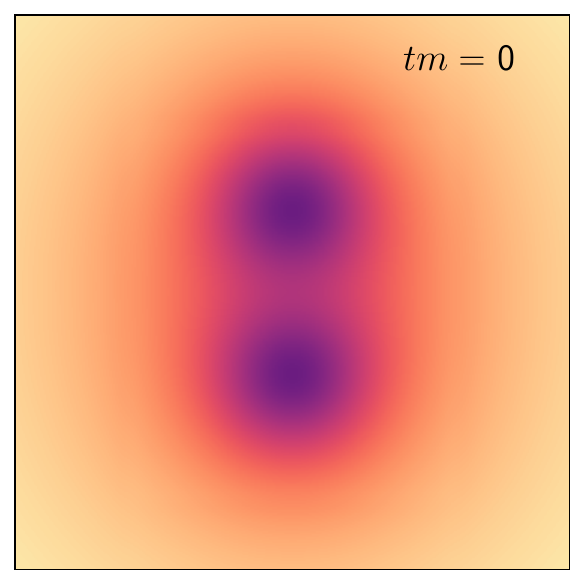}\hspace{-0.1cm}
  \includegraphics[height=0.13\textheight]{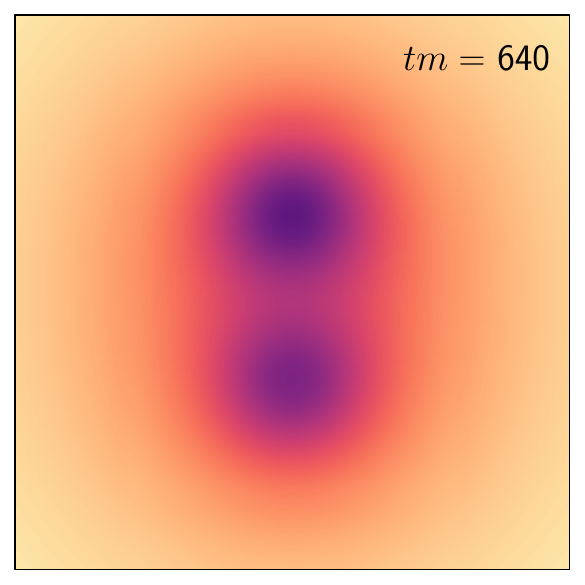}\hspace{-0.1cm}
  \includegraphics[height=0.13\textheight]{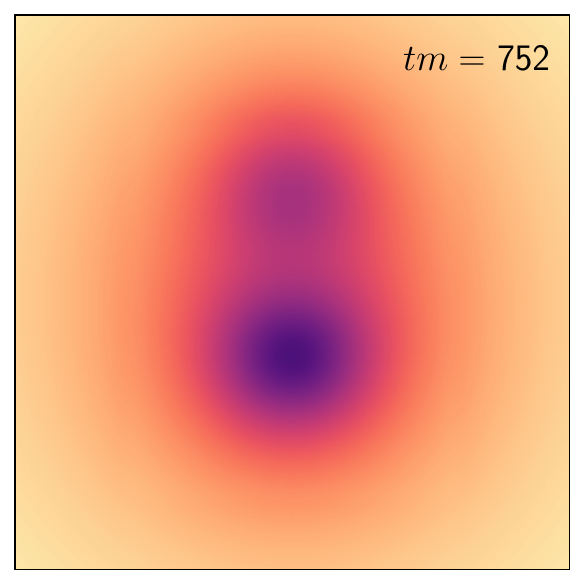}\hspace{-0.1cm}
  \includegraphics[height=0.13\textheight]{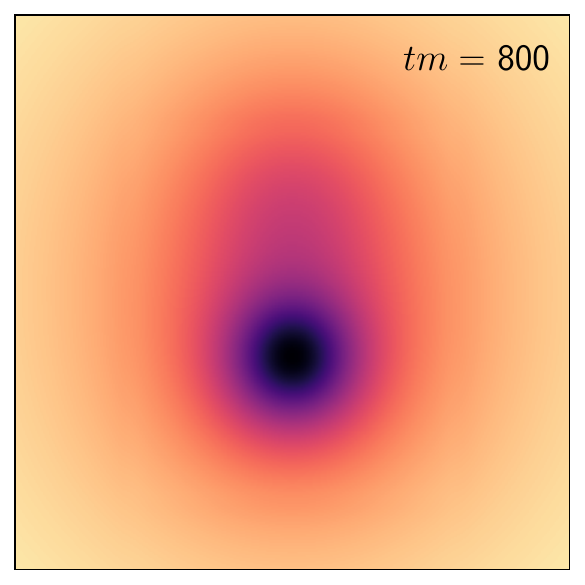}\hspace{-0.1cm}
  \includegraphics[height=0.13\textheight]{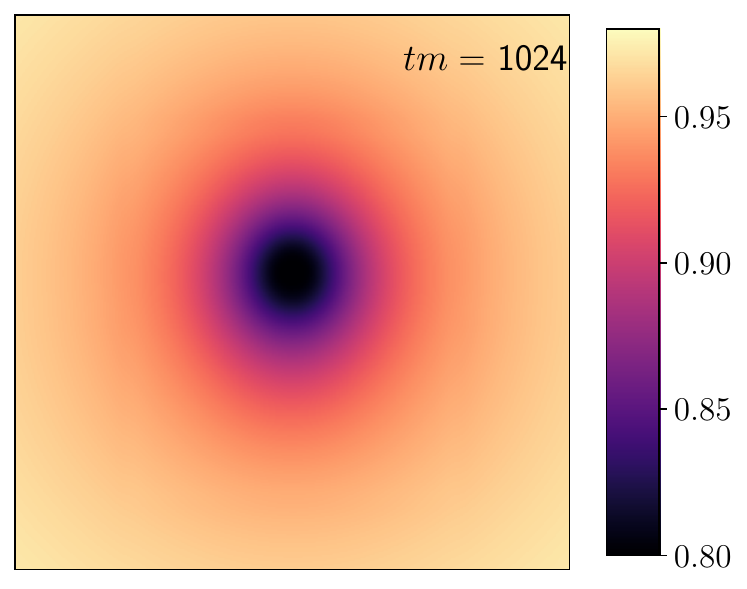}
  }
  \subfigure[~~Lapse function $N$]{
  \includegraphics[height=0.13\textheight]{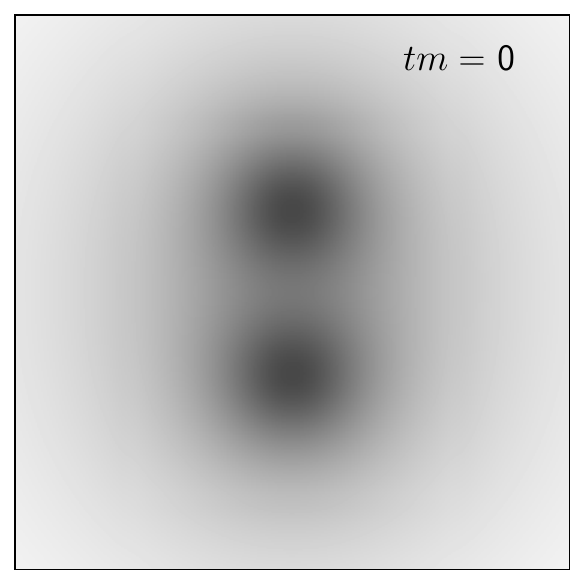}\hspace{-0.1cm}
  \includegraphics[height=0.13\textheight]{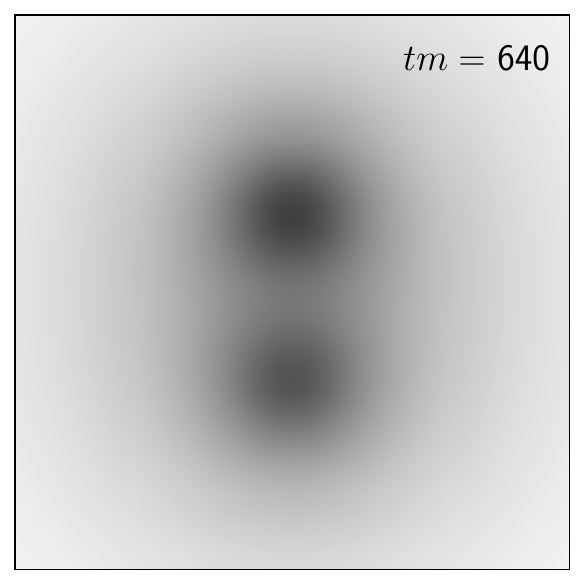}\hspace{-0.1cm}
  \includegraphics[height=0.13\textheight]{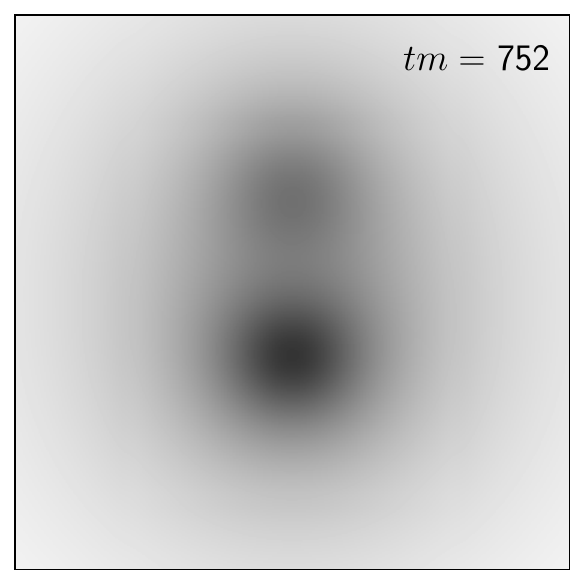}\hspace{-0.1cm}
  \includegraphics[height=0.13\textheight]{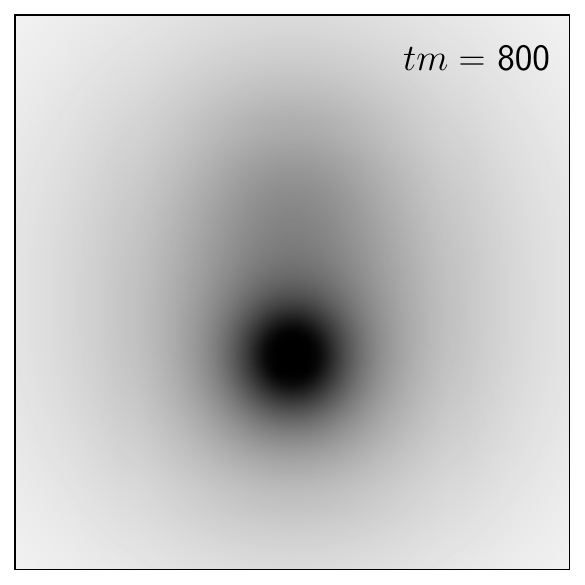}\hspace{-0.1cm}
  \includegraphics[height=0.13\textheight]{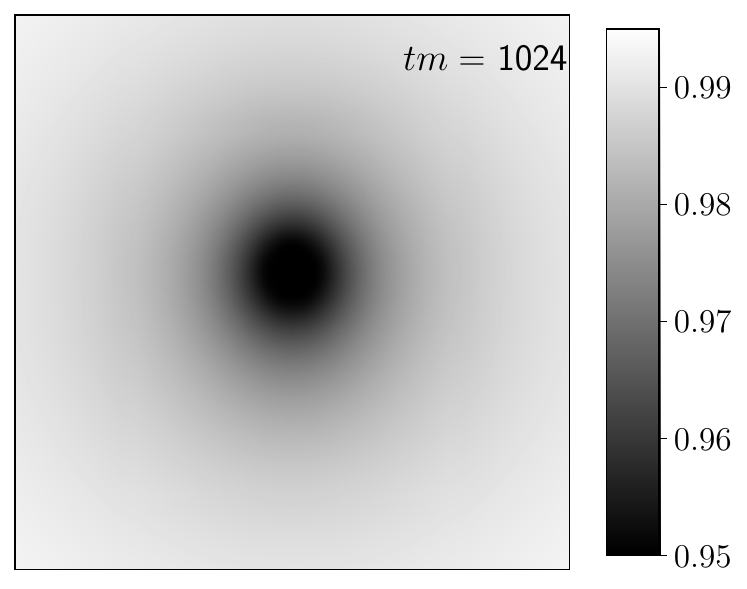}
  }
    \caption{
    A stable dipole with $\alpha^2 = 0.05$, $\mu=0.15$ and $\omega = 0.9$ (panel (a)) and development of the instability for the dipolar boson star with $\alpha^2 = 0.25$, $\mu=0$ and $\omega = 0.9$ (panels (b-d)). The unstable configuration correspond to a solution with initial separation $L=16.08$. The magnitude $|\Phi|$, the real field $\Psi$ and the lapse function of the 3+1 decomposition of the spacetime metric \cite{alcubierre2008introduction} are plotted in the $xz$ plane. The instability begins at $t=640$ and collapses to a spheroidal state around $t=800$. The configuration was slightly perturbed initially according to Eq.~\eqref{eq:perturbation}. The side of the box is 40 units in panel (a) and 50 units in panel (b).
    }
    \label{fig:dipole_destroyed}
\end{figure}

Figure~\ref{fig:dipole_destroyed} illustrates the typical behavior of the instability. Notably, in all cases studied, the instability remains axisymmetric, in contrast to the spinning case, where non-axisymmetric instabilities are present, as shown in \cite{Sanchis-Gual:2019ljs}. To verify the axisymmetric nature of the instability, we have compared the same quantities shown in Fig.~\ref{fig:dipole_destroyed} in both the $y=0$ plane (as presented) and the $x=0$ plane. Additionally, we verified the lapse function in the $z=0$ plane its axisymmetry throughout the evolution.

One important feature to note in Fig.~\ref{fig:dipole_destroyed}, particularly in panel (a), is that the growth of the perturbation is not symmetric with respect to the equatorial plane. Specifically, the energy density and spacetime metric exhibit a breaking of their even symmetry. The initial phase of the instability is characterized by an antisymmetric oscillation with respect to the $z=0$ plane, which intensifies over time and ends up in a perturbed spheroidal state. With this particular values of the parameters the end-point corresponds to a regular and horizonless solution however the instability developed in some other cases lead to the formation of a black hole. We will present some examples of this in Sec.~\ref{sec:Q-chains}

By further decreasing the value of $\alpha$ and focusing on the region $0 \leq \mu \leq 0.5$, we enter a parameter space zone where prominent peaks in the separation $L$ are observed, as seen in Fig.~\ref{fig:L}. In this region, we have identified stable solutions. To precisely delineate the stability region, we systematically explored the range $0.0125 \leq \alpha^2 \leq 0.25$ in increments of $\Delta \alpha^2 = 0.0125$, along with the case $\alpha^2 = 0.00625$. For each value of $\alpha$, we examined the range $0 \leq \mu \leq 0.25$ in steps of $\Delta \mu = 0.05$, as well as the case $\mu = 0.5$.

\begin{figure}
  \includegraphics[width=0.6\textwidth]{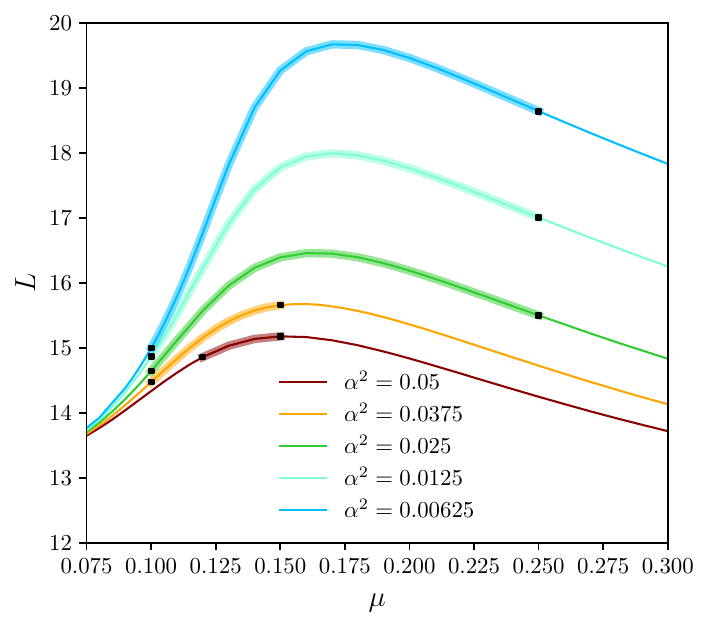}
    \caption{
    Stability region for dipolar  $\omega = 0.9$ boson stars and several cases of $\alpha$. The specific solution points were stability was found are displayed in Table~\ref{tab:stable}. For $\alpha^2 = 0.05$ we have performed an extra evolution with $\mu= 0.125$ (not shown in the Table).
    }
    \label{fig:stability}
\end{figure}

\begin{table}[b]
  \centering
  \begin{tabular}{lcc|cccc|c}
\hline
$\alpha^2$ & $\omega$ & $\mu$ & $L$ & $M$ & $Q$ & $\max(\Phi)$ & Result \\ 
\hline
0.05   & 0.9   & $\{0.15\}$                        & 15.18 & 0.226 & 0.229 & 0.332 & Stable \\
0.0375 & 0.9   & $\{\mathbf{0.1},0.15\}$           & 14.47 & 0.135 & 0.137 & 0.270 & Stable \\
0.025  & 0.9   & $\{0.1,0.15,\mathbf{0.2},0.25\}$  & 16.19 & 0.158 & 0.160 & 0.437 & Stable \\
0.0125 & 0.9   & $\{0.1,0.15,\mathbf{0.2},0.25\}$  & 17.76 & 0.0848&0.0855 & 0.462 & Stable \\
0.00625& 0.9   & $\{0.1,0.15,\mathbf{0.2},0.25\}$  & 19.46 & 0.0440&0.0443 & 0.474 & Stable \\
\hline
\end{tabular}
\caption{Stable dipolar FLS boson stars in the region $0\leq\mu\leq0.5$ for the cases of $\alpha$ presented in Fig.~\ref{fig:L}. In this region of the parameter space we have evolved all the $\alpha$ cases and varying the $\mu$ parameter in steps of $\Delta\mu=0.05$. Those configurations not present in the table are unstable. The quantities $L$, $M$ and $Q$ correspond to the solution with value of $\mu$ in bold of the corresponding row.}
\label{tab:stable}
\end{table}
\begin{figure}
  \includegraphics[height=0.18\textheight]{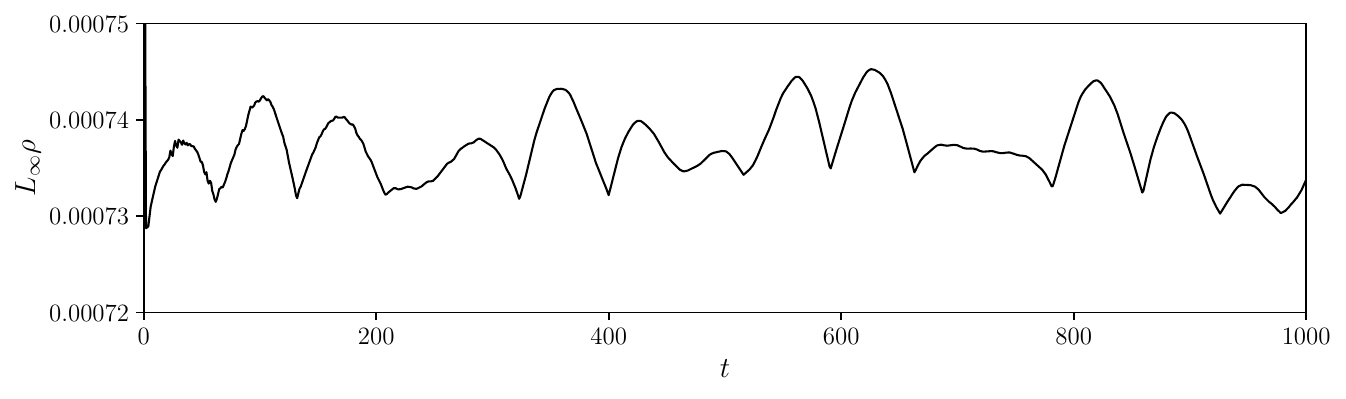}\\
  ~~\includegraphics[height=0.165\textheight]{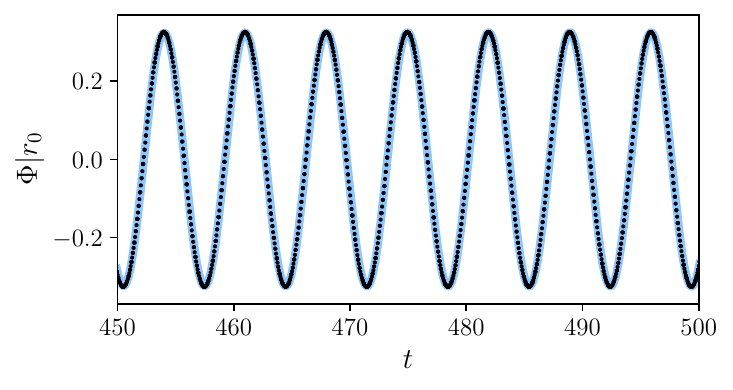}\hspace{-0.125cm}\includegraphics[height=0.165\textheight]{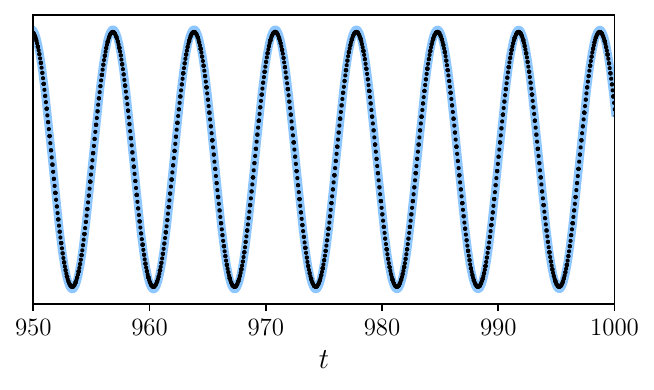}
    \caption{
      Stable dipole corresponding to the first row configuration in Table~\ref{tab:stable}. The configuration was perturbed according to Eq.~\eqref{eq:perturbation}. (Top panel) Time evolution of the maximum of $|\rho|$.  (Bottom panel) Time evolution of the real part of the scalar field extracted at the point $\mathbf{r}_0=(x=0,y=0,z=8)$ at the midpoint and the end of the simulation. For comparison we have added the plot of $\Phi|_{r_0,t=0}\cos(0.9t)$ to show that the scalar field continues to oscillate to a frequency very close to the initial one.
    }
    \label{fig:perturbed}
\end{figure}

From this extensive set of evolutions, we present in the Table~\ref{tab:stable} the stable configurations obtained. For clarity, Fig.~\ref{fig:stability} highlights the regions of stability. Among the unstable configurations in the range $0.0125 \leq \alpha^2 \leq 0.25$ and $0 \leq \mu \leq 0.25$, we observed that the timescale for the onset of instability can be quite long, extending up to $t = 5 \times 10^3$ in certain cases. So in order to further test the dynamical robustness of the dipoles that do not show signs of instability before $t = 10^4$, we introduced a forced perturbation at initial times which breaks the symmetry with respect to the equatorial plane in the $T_{\mu\nu}$ matter sources. This perturbation is applied as follows:
\begin{equation}\label{eq:perturbation}
\Phi(r,\theta;t=0) \to \Phi(r,\theta;t=0) + a \max(\Phi)\exp\left[-x^2-y^2-(z-L/2)^2\right] \, .
\end{equation}
The $\max(\Phi)$ correspond to the absolute maximum of the unperturbed $\Phi$ which always is obtained in $z>0$ and is reported in Table~\ref{tab:stable} and $a$ is a parameter that controls the strength of the perturbation.
This perturbation slightly increases the energy density near the ``star'' above $z = 0$, thereby breaking the symmetry along the $z=0$ plane. The resulting change in energy density requires an adjustment in the metric, which we achieve by solving the Hamiltonian constraint. Specifically, we apply the conformal thin sandwich formalism, solving the Lichnerowicz equation to obtain the new metric functions $F_1$ and $F_2$, which are conformally related to the original ones\footnote{Further details on the implementation can be found in \cite{Jaramillo:2024smx}.}.

We find that all the configurations listed in Table~\ref{tab:stable} remain stable, continuing to evolve around a dipolar static configuration close to the unperturbed state, even when subjected to the artificial perturbation with $a=0.05$. The perturbation applied is relatively strong, modifying in 5\% of the initial value of the complex field. Applying stronger perturbations drive some of the configurations, in particular those in the boundaries of the stability region represented in Fig.~\ref{fig:stability}, out of equilibrium.

To illustrate the resilience of these configurations, even under forced perturbations, in Fig.~\ref{fig:perturbed} we show the evolution of the highest $\alpha$ configuration found to be stable, corresponding to $\mu=0.15$, $\omega=0.9$, and $\alpha^2=0.05$ in Table~\ref{tab:stable}. The floor value of the energy density remains bounded and oscillating, while the harmonic dependence of the scalar field stays very close to the original frequency $\omega=0.9$ throughout the evolution. In these simulation we have verified that the geometrical quantities $N$ and $\gamma_{ij}$ follow the matter dynamics and also oscillate around a configuration close to the unperturbed state used in the preparation of the initial data.

\begin{figure}
  \includegraphics[height=0.18\textheight]{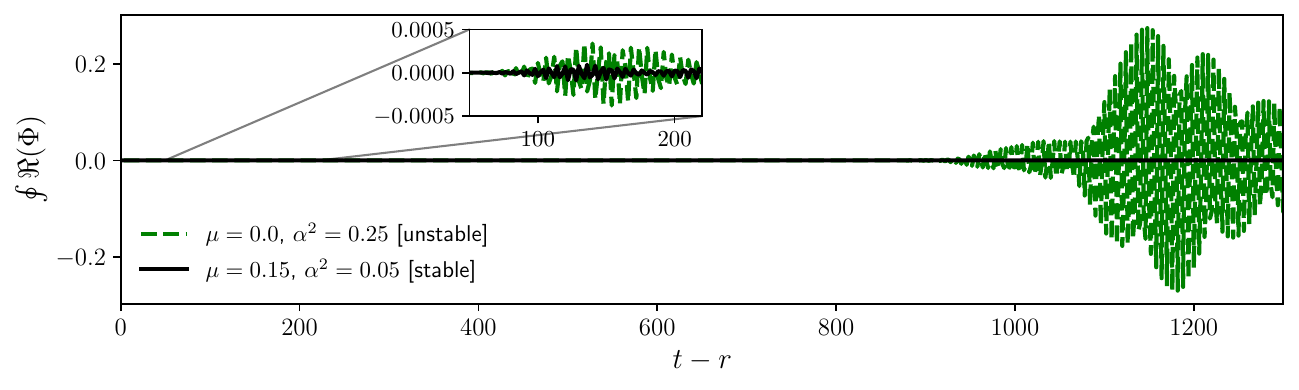}
    \caption{
      Surface integral of the real part of the scalar field $r\oint\Re(\Phi)d\Omega^2$ normalized by the maximum value of the unperturbed field at $t=0$, $\max(\Phi)$.
    }
    \label{fig:SW_grav_dipole}
\end{figure}

To further analyze the stability of these configurations, we examine the behavior of the scalar fields at a considerable distance from the source. For this purpose, we calculate the surface integral of the real part of the complex field $\Phi$ at a radius of 120, normalized by its maximum value at $t=0$. Fig.~\ref{fig:SW_grav_dipole} displays the scalar radiation produced by two configurations: an unstable one with $\mu=0$, $\alpha^2=0.25$ (as in Fig.~\ref{fig:dipole_destroyed}), and a stable configuration with $\mu=0.15$ and $\alpha^2=0.05$ (as in Fig.~\ref{fig:stability}).

In both cases, we apply a forced perturbation with amplitude $a=0.05$. As expected, an initial burst of radiation is seen around at initial times (see inset in Fig.~\ref{fig:SW_grav_dipole}). However, the $\mu=0.0$ configuration becomes unstable following this perturbation, leading to a second scalar field burst at $t \sim 1000$. No such instability is observed for the $\mu=0.15$ configuration, which remains stable throughout the evolution.

Although the configurations with $\mu = 0$ and $\mu = 0.05$ fall outside the region of stability and may not exhibit significant dynamical robustness, examining self-gravitating solutions in this regime provides important insights. As the value of $\alpha$ decreases, the separation $L$ and other related quantities remain well-behaved. This observation strongly suggests that such solutions can persist in the flat spacetime limit, offering clues for the potential construction of non-rotating Q-ball chains.

\section{Chains}\label{sec:Q-chains}

Up to this point, we have focused on the simplest type of static parity-odd solutions within the Einstein-FLS system. These solutions involve a spacetime metric $g_{\mu\nu}$ and a real field, $\Psi$, with even parity relative to the equatorial plane, while the scalar field, $\Phi$, exhibits odd parity. In these cases, $|\Phi|$, has a single node along the symmetry axis, specifically at $z=0$. Now, we will explore more complex still axisymmetric configurations where $\Phi$ remains parity-odd, but $|\Phi|$ presents certain number of nodes, $k_z$, along the symmetry axis. In particular we will consider $k_z=1$, 3 and 5, corresponding to two component, four component and six component chains. These structures can be interpreted as chains formed by an even number of constituents. We begin by constructing these configurations in flat space, where Q-chains are more straightforward to generate by combining dipolar setups. Once the chains are established, we will gradually increase $\alpha$, allowing the configurations to become gravitating chains of boson stars.

\subsection{Q-dipoles and chains in flat space}

Multisoliton solutions in flat spacetime can be constructed by starting with a self-gravitating dipole and gradually decreasing the value of $\alpha$ until it reaches zero. Given a pair of parameters $(\mu, \omega)$, we can investigate the existence of such solutions using this approach. We began our search with the configurations listed in the broad scan of Table~\ref{tab:dipolar} and found $\alpha=0$ solutions only for the case $\mu=0$. This outcome is expected, as seen in Fig.~\ref{fig:L}, where, among the cases $\mu=0$, 0.25, and 0.5, only $\mu=0$ shows that the separation $L$ does not grow indefinitely as $\alpha$ approaches zero.

We calculated the separation $L$ while varying $\alpha$ for configurations with $\omega=0.8$ and $\omega=0.7$, finding that the region near $\mu=0$, where flat space solutions are expected to exist, expands as $\omega$ decreases. For example for the case $\omega = 0.9$, it was possible to construct solutions up to a value $\mu=0.1379$, after this point the total energy, change and separation of the solutions increase more steeply and the code is not able to converge to solutions with bigger values of $\mu$ when considering steps in the construction proceedure of $\Delta\mu=0.0001$. For the case $\omega=0.8$ we find that the maximum value of $\mu$ allowed by our construction procedure is $\mu=0.1964$ while for the $\omega = 0.7$ is $\mu=1.200$.

\begin{figure}
  \includegraphics[width=0.46\textwidth]{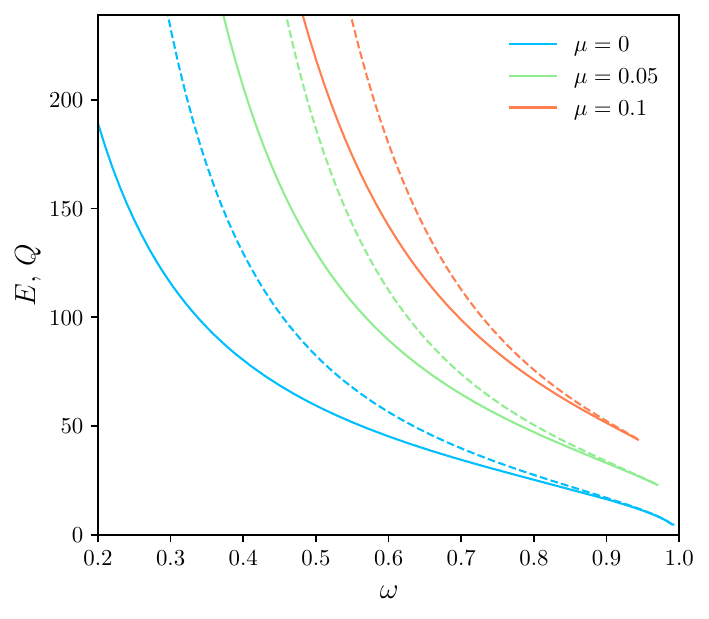}
  \includegraphics[width=0.45\textwidth]{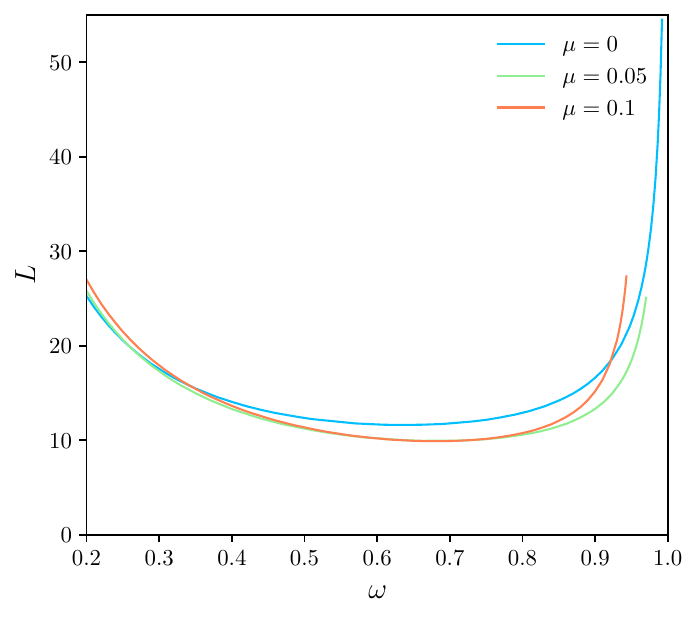}
    \caption{
      Two static Q-balls in equilibrium. Sequence of solutions for $\alpha = 0$ and $\mu = \{0,0.05,0.1\}$. In the left panel continuous lines represent $E$ and dashed lines $Q$.
    }
  \label{fig:flat}
\end{figure}

Based on this analysis, for a given $\mu$, dipolar solutions in flat space exist and can be constructed for certain range of the frequency with a maximum value of $\omega_{\rm max}$ which decreases as $\mu$ increases. We confirmed this by constructing sequences of dipolar Q-balls for three different values of $\mu$. The resulting charge, energy 
\begin{equation}\label{eq:energy}
  E=\int\rho \sqrt{-g} d^3x \, = M ~~~~ \text{(flat spacetime)},
\end{equation}
 and separation are presented in Fig.~\ref{fig:flat}.

The energy/charge-frequency diagram for the $\mu=0$ case bears a qualitative resemblance to the corresponding monopolar, spherically symmetric FLS Q-ball solutions with $\mu=0$ (see Fig.~2 in \cite{Loiko:2018mhb}). However, there is a key distinction: for solutions approaching the $\omega=1$ limit, the separation between the two lumps appears to diverge. At this limit, we observe a significant increase in the numerical error indicator—particularly the virial identity \cite{Herdeiro:2021teo,Shnir:2018yzp,Herdeiro:2023lze} used to establish reliability of the flat spacetime solutions—which signals a loss of accuracy. As a result, solutions with higher values of $\omega$ must be discarded. Finally we notice that all configurations are classically stable (the energy of the free scalar quanta is bigger than the energy of the system $Q>E$) in the three sequences constructed.

Now we explore the possibility to build richer structure of static Q-balls in axisymmetry. In particular we proceed to construct \textit{even chains} of Q-balls. This kind of solutions \cite{Loiko:2020htk,Herdeiro:2021mol,Sun:2022duv} can be interpreted as an even number of Q-balls located along the symmetry axis and equilibrated (as the dipolar case) by the attractive scalar interactions and the repulsion coming from the out of phase interaction between components. Beyond the dipolar chain already presented, we will present four and six constituents solutions.

To construct these configurations and ensure a sufficiently accurate initial guess, we begin with one of the previously obtained dipolar solutions, $\Phi_{d}$. We then combine multiple copies of this field, maintaining the odd parity for the four and six-component trial functions, respectively:
\begin{eqnarray}
  \Phi_q &=& \Phi_d(x,y,z-l) + \Phi_d(x,y,z+l) \, ,\\
  \Phi_h &=& \Phi_d(x,y,z-2l) + \Phi_d(x,y,z) + \Phi_d(x,y,z+2l) \, ,
\end{eqnarray}
where $l$ is a value close to the separation $L$. For the real field, we initially set $\Psi = 1$. Starting from this initial guess, the solver successfully converges to a solution.

\begin{figure}
  \subfigure[~~$\Phi$]{
  \includegraphics[height=0.15\textheight, angle=0]{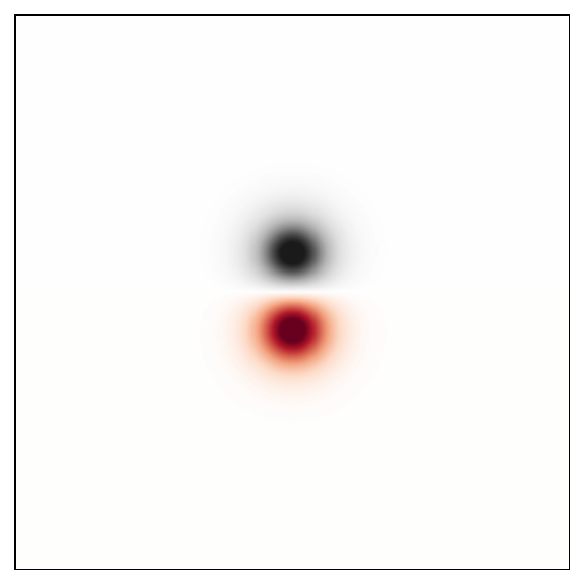}
  \includegraphics[height=0.15\textheight, angle=0]{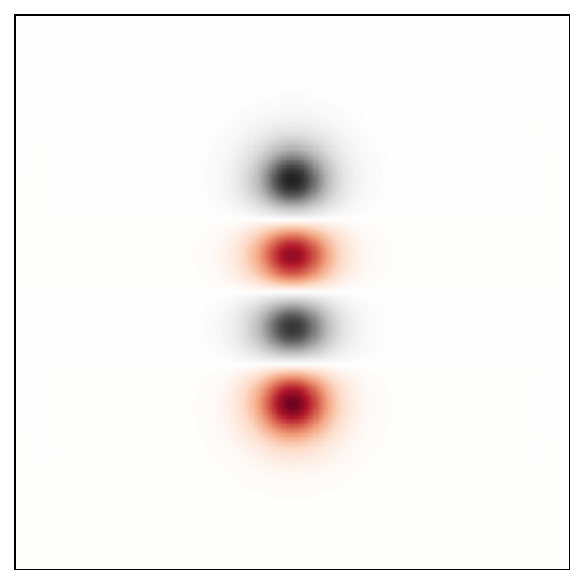}
  \includegraphics[height=0.15\textheight, angle=0]{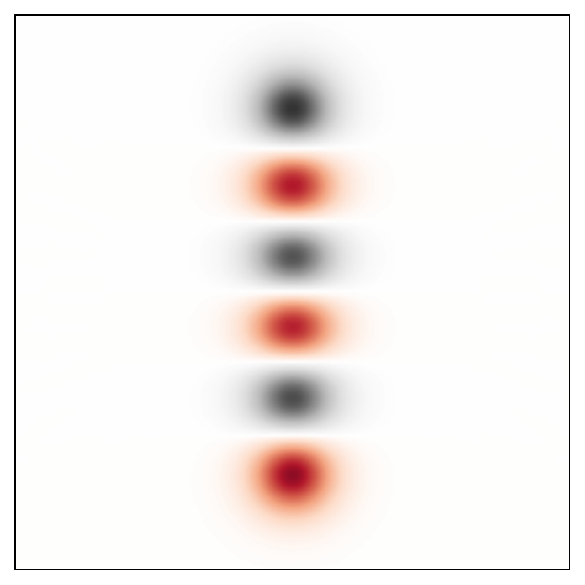}}\\
  \subfigure[~~$\Psi$]{
  \includegraphics[height=0.15\textheight, angle=0]{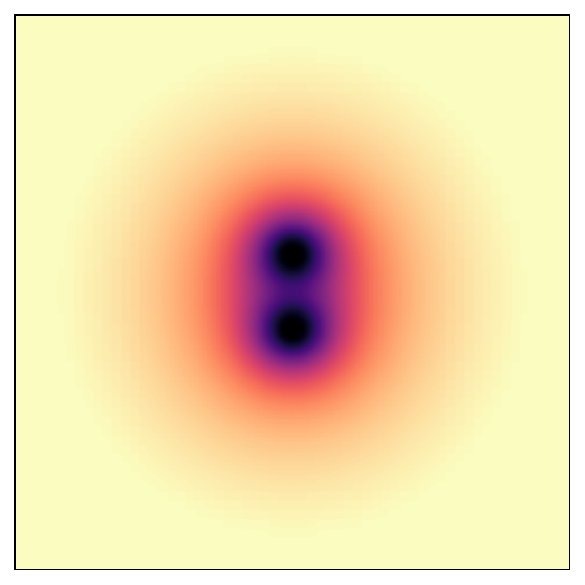}\hspace{0.035cm}
  \includegraphics[height=0.15\textheight, angle=0]{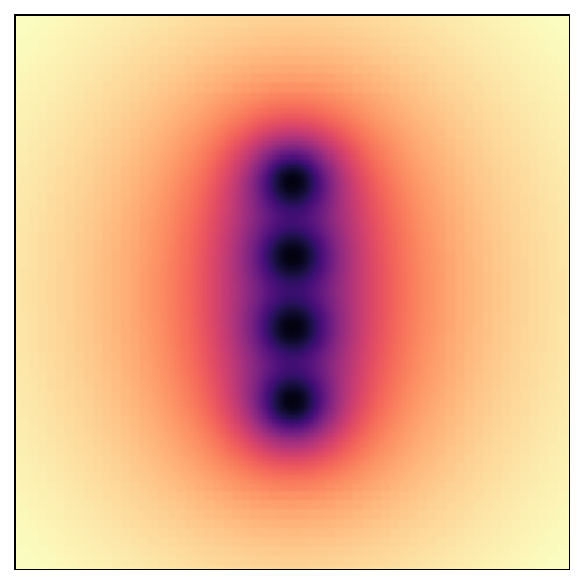}\hspace{0.035cm}
  \includegraphics[height=0.15\textheight, angle=0]{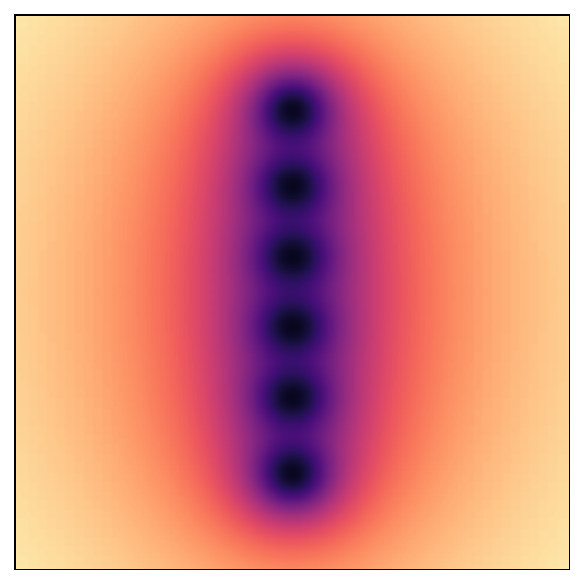}}
    \caption{
      Q-chains with two, four and six components ($k_z=1$, 3 and 5) for the massless real scalar field $\mu=0$, and frequency $\omega=0.9$. In the plots we use coordinates $\rho= r\sin \theta$ and $z = r\cos \theta$,
      The vertical direction is the $z$-axis, while
      the horizontal direction is the $\rho$-axis. The real field $\Psi$ is long-ranged in this limit. The side of the box is 120.
    }
  \label{fig:chains}
\end{figure}

Examples of these chain solutions are shown in Fig.~\ref{fig:chains}. To our knowledge, no other non-rotating scalar field chains in flat spacetime involving Q-balls have been reported in the literature. However, rotating solutions with odd parity in the Maxwell-FLS model were constructed in \cite{Loiko:2020htk}. Also odd and even parity spinning chains for self-interacting scalar fields, have been documented \cite{Gervalle:2022fze}; this reference also notes challenges in constructing non-rotating Q-chains within self-interacting scalar field theory, specifically an increase in distance between neighboring constituents as $\alpha \to 0$. Additionally, the existence of odd-parity self-interacting solutions in flat spacetime was ruled out in \cite{Cunha:2022tvk} using mathematical arguments. Therefore, since Q-balls are absent in the simple free field case and dipoles or Q-chains seem to be also absent in the self-interacting case, FLS configurations appear to be the simplest type of non-topological chains constructed in 3D so far.

The solutions displayed in Fig.~\ref{fig:chains} were obtained by fixing the parameter $\omega = 0.9$. While previous configurations were generated using a decomposition with 17 spectral coefficients in both $r$ and $\theta$, the six-component solutions required 25 coefficients, to achieve the same level of precision as the dipolar cases. It can be observed from Fig.~\ref{fig:chains} that the overall size of the configuration along the symmetry axis increases as more components are added. However, the spacing between individual lumps and their amplitudes remains relatively consistent. Consequently, for a fixed $\omega$, we expect the four-component chain to have twice the mass of the dipole, while the six-component configuration should be three times the mass of the dipole. This pattern suggests that it is feasible to continue stacking dipoles to construct higher-order chains.

\begin{figure}
  \includegraphics[width=0.49\textwidth]{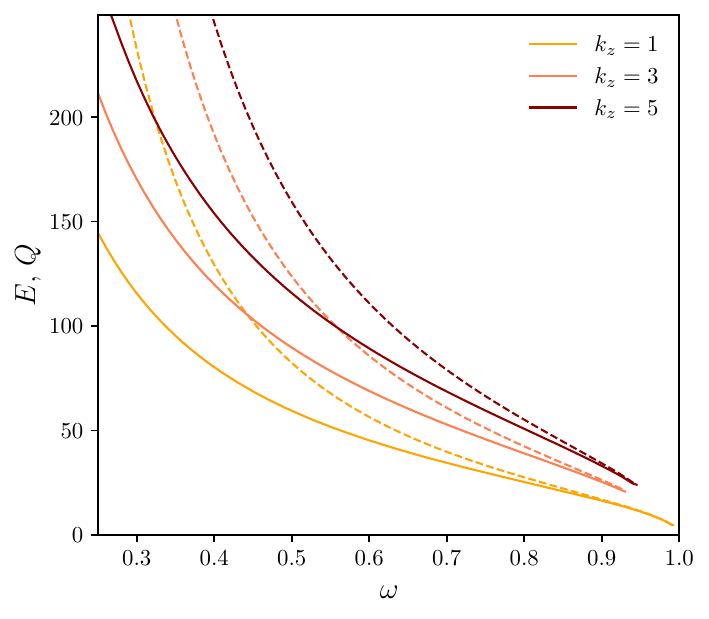}
    \caption{
      Parity-odd Q-chains. Sequence of equilibrium solutions for $\mu = 0$ for the two, four and six component chains. The solutions are classically stable
    }
  \label{fig:Mchains}
\end{figure}

The solutions in Fig.~\ref{fig:chains} were used to build the $k_z=3$ and $k_z=5$ sequence of solutions for the case $\mu=0$ by varying the value of the frequency $\omega$. We present the energy/charge-frequency diagram of such families in Fig.~\ref{fig:Mchains}. Solutions also exist closer to $\omega=1$, converging towards configurations of widely separated, nearly isolated Q-balls in equilibrium.
As seen in Fig.~\ref{fig:Mchains}, similar to the case of spherical flat Q-balls with a vanishing potential $\mu=0$ \cite{Loiko:2018mhb}, there is only a single branch of solutions. Throughout this branch, we observe that $Q > E$, suggesting that these solutions are potential candidates for stability.

We tested the stability of the three configurations with $\omega = 0.9$ shown in Fig.~\ref{fig:Mchains}. To do this, we evolved the initial data corresponding to these equilibrium solutions using Eqs.~\eqref{eq:bssn_fields} within the same numerical framework described in Sec.~\ref{sec:setup}. However, we set $\alpha = 0$ and disabled the evolution of the geometric sector of the Einstein equations. We then conducted a stability analysis similar to the one described in Sec.\ref{sec:evolutions}.

\begin{figure}
  \includegraphics[height=0.13\textheight]{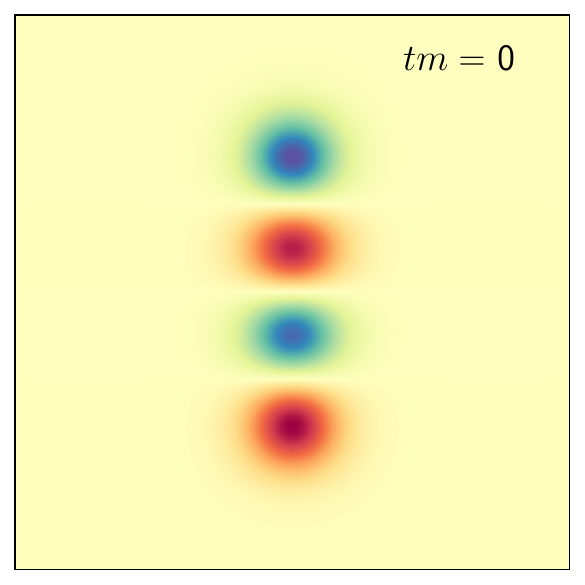}\hspace{-0.1cm}
  \includegraphics[height=0.13\textheight]{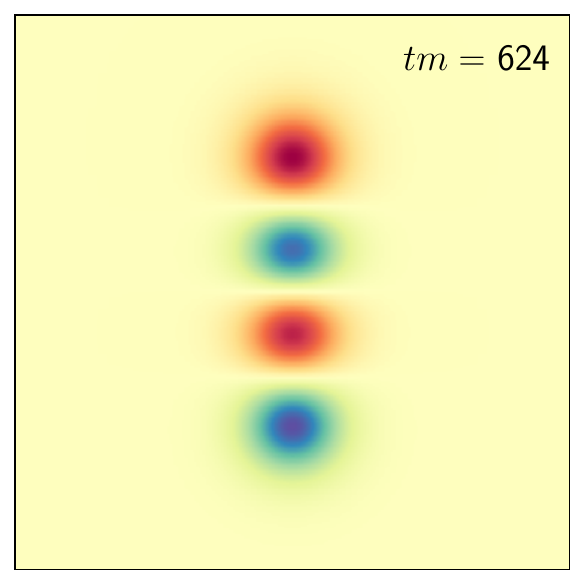}\hspace{-0.1cm}
  \includegraphics[height=0.13\textheight]{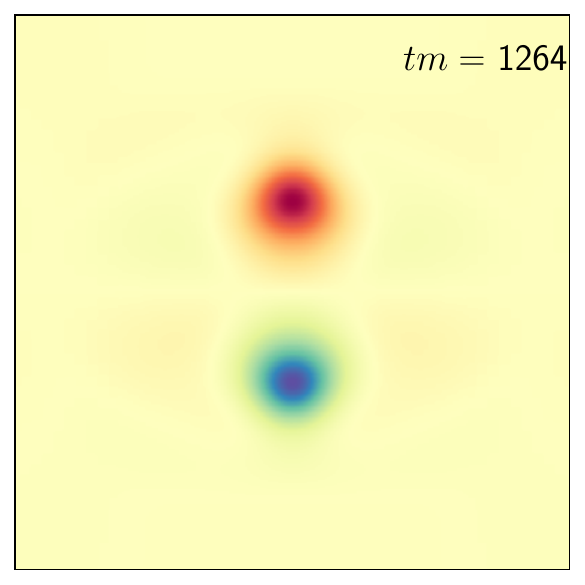}\hspace{-0.1cm}
  \includegraphics[height=0.13\textheight]{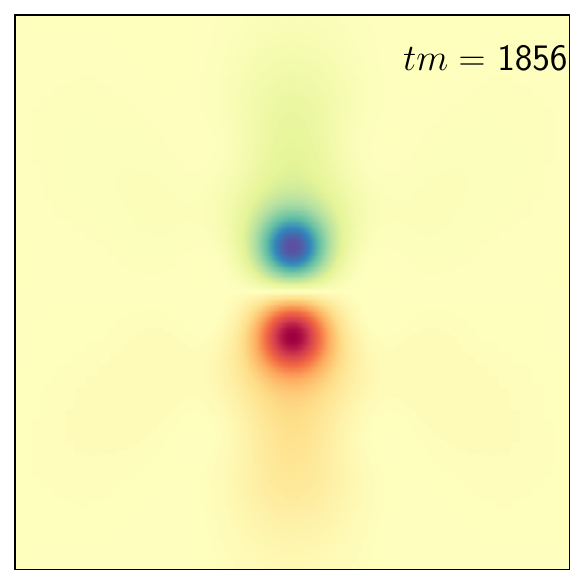}\hspace{-0.1cm}
  \includegraphics[height=0.13\textheight]{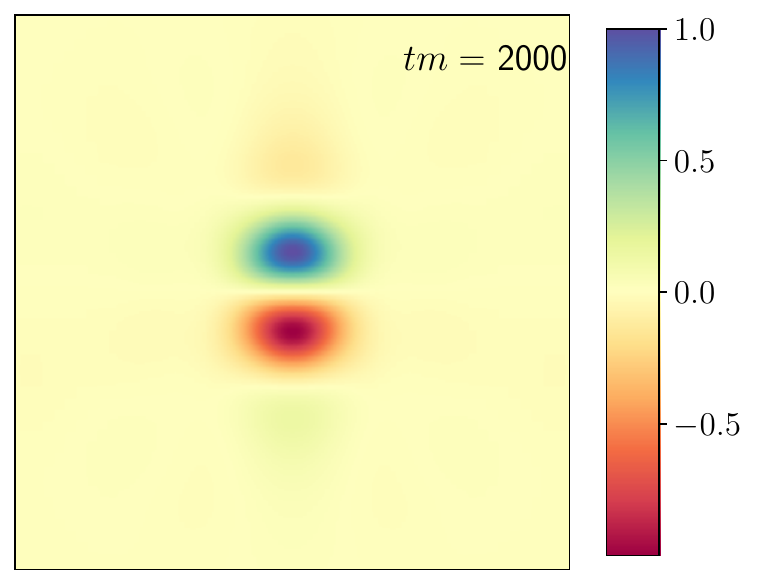}
    \caption{
    Development of the instability for the unperturbed $k_z=3$ chain with $\omega = 0.9$ and $\mu = 0$. The plot shows the normalized real part of the complex scalar field, $\Phi_r / \max(|\Phi_r|)$, in the  $xz$-plane. The instability initiates at $t = 624$  and evolves into a highly dynamic dipolar configuration around $t ~ 1000$. The side length of the box measures 100 units.
    }
    \label{fig:chain_destroyed}
\end{figure}

The results indicate that all three configurations—the $k_z=1$, 3 and 5 chains with $\mu = 0$ and $\omega = 0.9$—are unstable, decaying before $t = 2000$ into simpler morphologies under the influence of numerical truncation errors.\footnote{
To evolve the chains, we had to increase the size of the refinement levels and adjust their shape until energy conservation was observed in the code. Due to the computational cost associated with evolving these larger boson star systems, the unperturbed configurations were evolved only up to $t = 2000$.
} 
For instance, the two-component chain collapses into a spheroidal monopolar dynamic state, while the four-component chain transitions into a clear dipolar configuration (see Fig.~\ref{fig:chain_destroyed}). Notably, the four and six-component configurations maintain symmetry with respect to the equatorial plane during their evolution. However, due to the limited evolution time considered, it remains unclear whether these configurations will eventually settle into any of the corresponding lower-component equilibrium states, as no definitive equilibrium has been reached.

Based on the results from the self-gravitating dipole, where none of the $\mu = 0$ configurations were found to be stable, but a region with small non-zero $\mu$ demonstrated stability, we have also constructed configurations with  $\mu > 0$. Specifically, we examined cases with $\mu = 0.05$, 0.1 and 0.125. As discussed earlier, the maximum value of $\mu$ where flat space dipolar solutions were obtained was $\mu = 0.1379$; hence, we anticipate a similar upper limit for the $k_z > 1$ chains.

\begin{figure}
  \subfigure[]{
  \includegraphics[height=0.2\textheight]{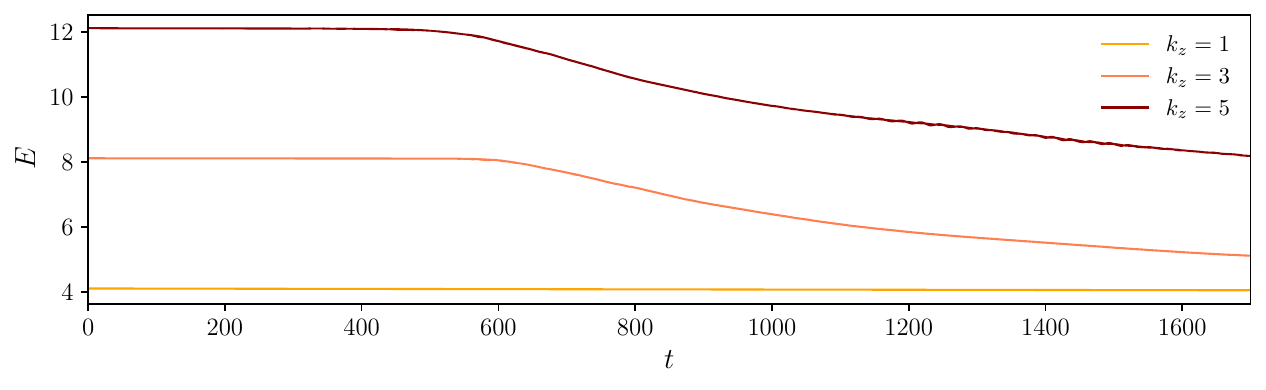}}\\
  \subfigure[]{
  \includegraphics[height=0.2\textheight]{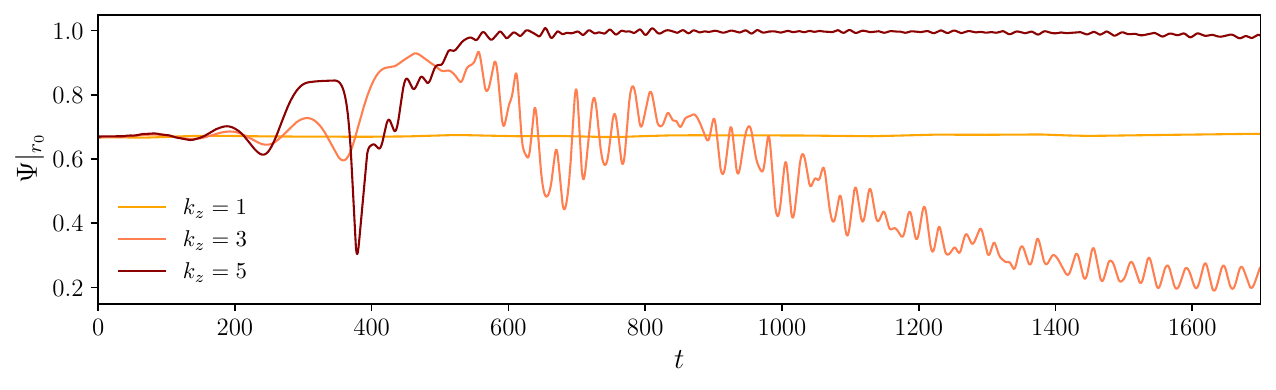}}
    \caption{
      Evolutions of two, four, and six component chains with $\mu = 0.1$ and $\omega = 0.9$. The initial configurations were perturbed according to Eq.~\eqref{eq:perturbation}, with $L$ representing the distance between the two innermost components of the chain. In panel (a), the total energy \eqref{eq:energy} of the chains is shown, while panel (b) displays the time evolution of the real scalar field $\Psi$, measured at the point $\mathbf{r}_0 = (x = 0, y = 0, z = 8)$. The dipolar configuration is stable.
    }
    \label{fig:perturbedQ}
\end{figure}

Our findings show that configurations with $\mu > 0.05$ survive significantly longer when initially unperturbed. However, applying an artificial perturbation that explicitly breaks the symmetry of $|\Phi|$ with respect to the equatorial plane—similar to the perturbation in Eq.~\eqref{eq:perturbation} destroys the $k_z=3$ and $k_z=5$ Q-chains. These chains develop a non-symmetric perturbation and eventually decay into a monopolar state. In Fig.~\ref{fig:perturbedQ}, we illustrate the onset of this instability for $\mu = 0.1$ happening for the $k_z = 3$ and $k_z = 5$ cases but not for $k_z=1$.
We have also tested the case with $\mu = 0.125$, which, based on our tests, is stable for the dipolar configuration. However, we found that the four and six component chains with this value of $\mu$ are still unstable. Nevertheless, we do not rule out the possibility that other chains with $k_z > 1$  might exhibit stability, especially when considering less massive systems. A more comprehensive study on the stability of such chains will be presented in future work.

\subsection{Gravitating chains}

Solitons in a free, massive complex scalar field cannot exist in flat spacetime but can emerge when coupled to gravity. Gravity effectively regularizes both spherical solutions and chains, which would otherwise diverge at the origin in the linear Klein-Gordon equation context \cite{Herdeiro:2020kvf}. In previous sections, we have demonstrated that coupling the complex massive scalar field to a real scalar field within the FLS model allows solutions to be constructed in flat space, provided the real scalar field is long-ranged (near $\mu=0$). The equilibrium resulting from the long-range gravitational attraction can thus be mimicked by the long-range interaction introduced through the real scalar field. In the following paragraphs, we show that self-gravitating even chains in the Einstein-FLS model exist and are smoothly connected to the flat spacetime solutions as the parameter $\alpha$ is turned on.

We begin by fixing the frequency at $\omega = 0.9$, located within the fundamental branch (between the trivial $M=0$ limit and the global minimum $\omega$ solution), and gradually increase the value of $\alpha$ for four and six component chains with $\mu = 0$. We conjecture that, similar to $k_z=1$ solutions, chains with $k_z>1$ will also permit a broader range of $\mu$ values as $\alpha$ increases. In line with this, we incremented $\alpha$ up to $\alpha^2=0.25$. According to the analysis in Sec.~\ref{sec:static}, configurations with this relatively large $\alpha$ resemble certain global properties of dipolar boson EKG stars, but with notable differences from the effect of the FLS parameter $\mu$ as $\mu$ approaches small values in the range $\mu \in [0, 0.5]$ (see, e.g., Figs.~\ref{fig:L}, \ref{fig:M}, and \ref{fig:Q}). We obtain that in the region $\alpha^2 \in [0,0.25]$, both $M$ and $Q$ decrease slightly and almost linearly, as does the separation $L$ between the innermost peaks of $|\Phi|$. This behavior reflects the attractive nature introduced by the additional interaction, which tends to contract the configuration, thereby accommodating less matter within the chain of stars.

\begin{figure}
  \includegraphics[width=0.48\textwidth]{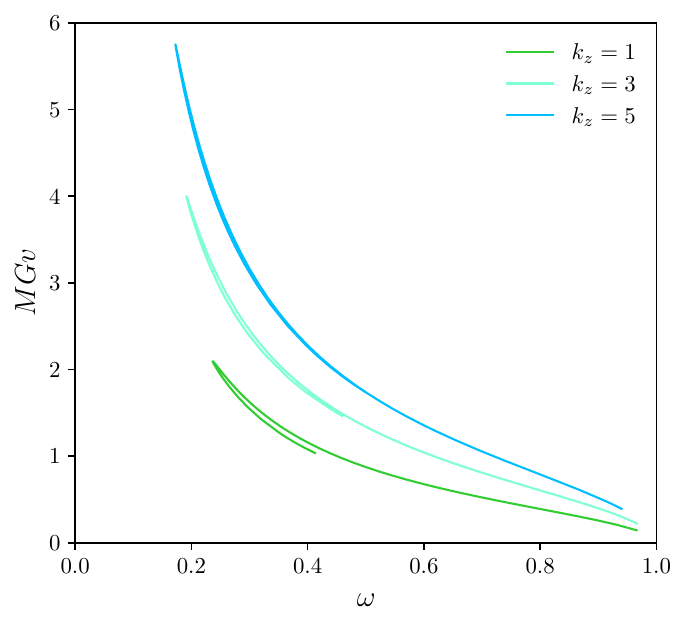} \includegraphics[width=0.49\textwidth]{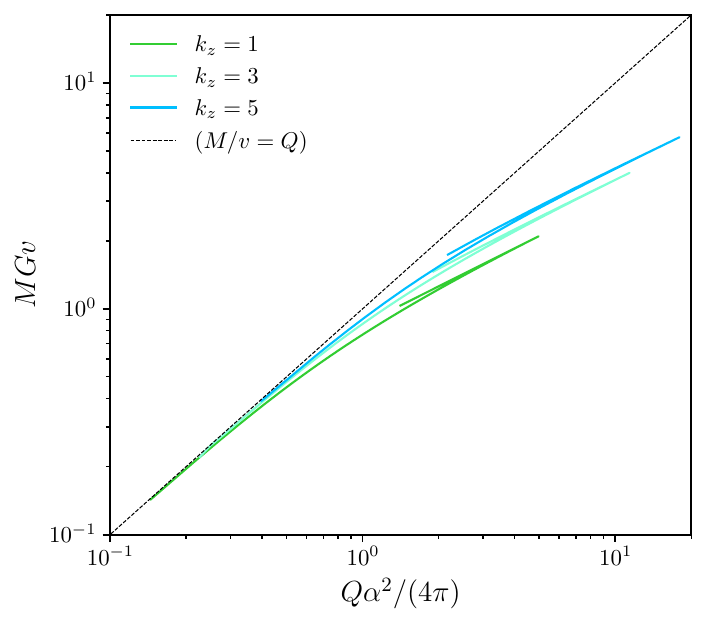}
    \caption{
    Gravitating chains with two, four and six components for $\mu=0$ and $\alpha^2 = 0.25$. (Left panel) Mass-frequency diagram. (Right panel) Energy as a function of the Noether charge for the chains.
    }
    \label{fig:MQ-grav_chains}
\end{figure}

In the left panel of Fig.~\ref{fig:MQ-grav_chains}, we present the total mass $M$ for chains with $\alpha^2=0.25$ and $\mu=0$ in the $k_z=3$ and $k_z=5$ cases as a function of the frequency $\omega$. For comparison, the $k_z=1$ dipole, already shown in Fig.~\ref{fig:M}, is included. As with dipoles, the range of $\omega$ values over which equilibrium solutions exist is broad and even expands slightly for larger $k_z$. As expected, the mass of higher-order chains also increases with $k_z$. In the right panel of Fig.~\ref{fig:MQ-grav_chains}, we show a mass-charge diagram for these solutions, indicating that, like dipoles, these chains appear to be gravitationally bound throughout their entire existence domain.

\begin{figure}
  \includegraphics[width=0.48\textwidth]{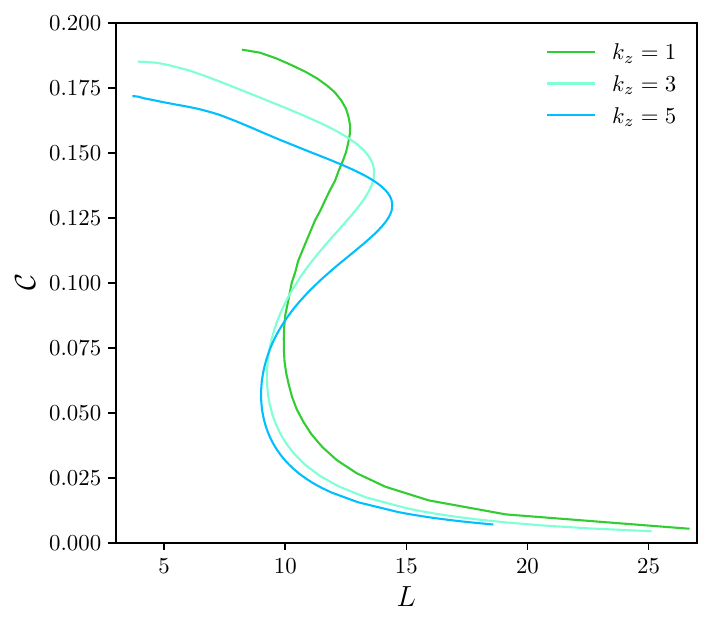} \includegraphics[width=0.475\textwidth]{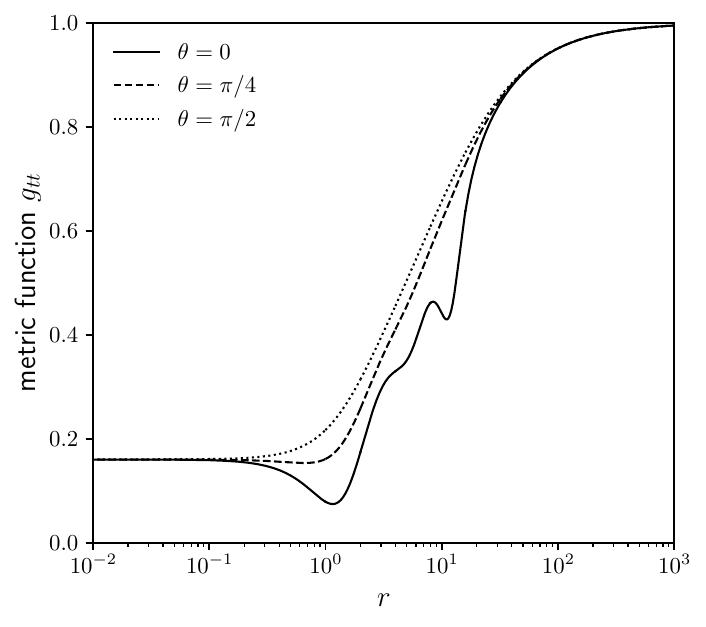}
    \caption{
    Compactness of gravitating chains for $\mu=0$ and $\alpha^2 = 0.25$. (Left panel) Compactness vs. the separation $L$ between the central nodes of $|\Phi|$. (Right panel) Metric function $-g_{tt}=N^2$ as a function of $r$ for three different angles $\theta$ and for the configuration in the second branch with $\omega = 0.366$ and $k_z=5$, the chains are located along $\theta=0$.
    }
    \label{fig:C-grav_chains}
\end{figure}

An important observation is that, unlike the flat space chains (e.g., Fig.~\ref{fig:Mchains}), the mass of these gravitating configurations does not diverge as they approach the minimum allowed frequency. Instead, they follow the characteristic spiral pattern in $M(\omega)$, where the configurations become increasingly compact. In these constructed chains, we have achieved solutions within the second solution branch (beyond the fundamental branch shown in Fig.\ref{fig:MQ-grav_chains}). Extending further along this branch is possible; however, as configurations become more compact in this region, precise tuning of solver parameters is necessary for continued solutions. To demonstrate this trend toward compactness along the second branch, we show a measure of the ``compactness'' in Fig.~\ref{fig:C-grav_chains}. Here, compactness is defined as the total mass of the chain divided by $k_z L$, with $L$  representing the proper distance between the two closest nodes of $|\Phi|$ along the symmetry axis.
\begin{equation}
  \mathcal{C} = \frac{M}{k_z L} \, .
\end{equation} 

Alongside the compactness of the chains, we have plotted the lapse function for a configuration with $k_z = 5$  in the right panel of Fig.~\ref{fig:C-grav_chains}. We observe that the metric function $g_{tt}$ is close to zero in some region, where based on the static line element \eqref{eq:metrica} would imply the presence of and event horizon. We evolved this configuration, which is expected to be unstable, and confirmed that it collapses to form a black hole after some time.

To carry out this evolution, we adapted the previous setup, increasing the refinement levels from three to seven, and used a physical grid with a side length of 400 units, twice the size of previous cases. This setup required significantly higher resolution due to the configuration’s smaller size, stronger fields and gradients of the fields, and the need for a larger domain to capture the emitted gravitational waves.

The gravitational wave emission was calculated by decomposing the Newman-Penrose scalar $\Psi_4$ into spin-weighted spherical harmonics with $s = -2$,
\begin{eqnarray}
  \Psi_4(t,r,\vartheta,\varphi) = \sum_{l, m}\Psi_4^{l,m}(t,r) \, {}_{-2}Y_{l,m}(\vartheta,\varphi) \ .
\end{eqnarray}
Far from the source, $\Psi_4$ serves as a measure of gravitational radiation during the merger of compact objects, relating to the gravitational wave strain $h = h_+ - i h_\times$  by $\Psi_4 = -\partial_t^2 h$ and decaying as $1/r$ \cite{Wald:1984rg}. To extract the gravitational wave signal, we computed the $\Psi_4^{l, m}$ coefficients for $l = 2$  and  $m = 0$, 1, 2 at radii $r = 80$, 100, 120, and 140, using the \texttt{Multipole} and \texttt{WeylScal4} thorns within the toolkit used for the evolutions. We confirmed the expected $1/r$ decay and report only the coefficients extracted at $r = 140$.

\begin{figure}
  \includegraphics[width=0.82\textwidth]{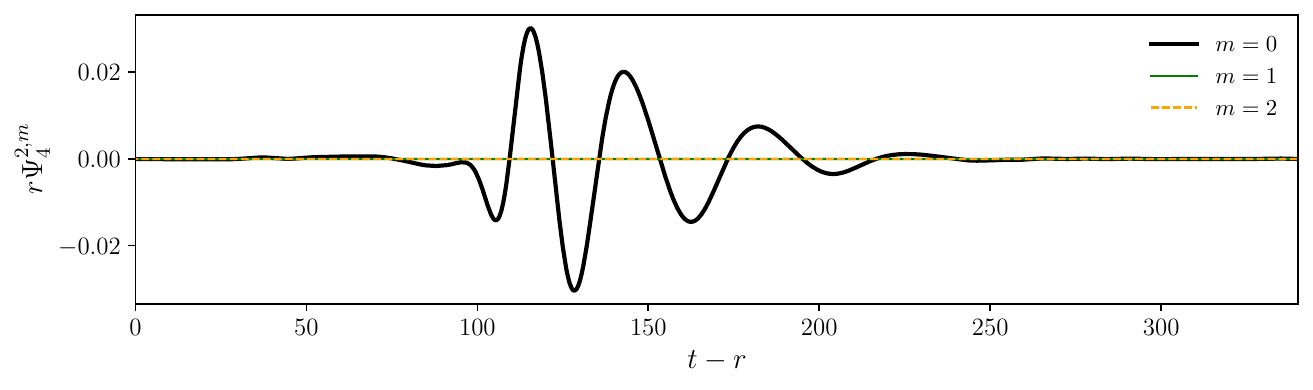} \\
  \hspace{0.3cm}\includegraphics[width=0.8\textwidth]{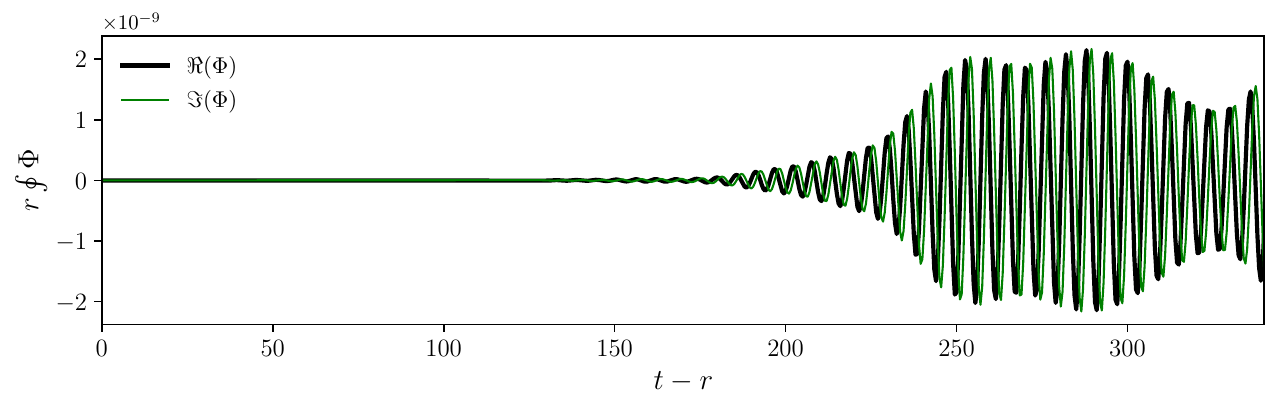} 
    \caption{
      Gravitational and scalar wave of a the collapsing 6 component chain with $\alpha^2 = 0.25$, $\mu=0$ and $\omega = 0.366$ in the second branch. This is the same configuration presented in the right panel of Fig.~\ref{fig:C-grav_chains}. (Top panel) Real part of the $l=2$ components of $\Psi_4$ rescaled in both axis according to the extraction point $r=140$. (Bottom panel) Surface integral of $\Phi$ at the sphere of radius $r=140$. 
    }
    \label{fig:GW-grav_chains}
\end{figure}

In Fig.~\ref{fig:GW-grav_chains}, we present the real part of $r\Psi_4^{2,0}$ (the dominant mode due to the symmetry of the configuration and subsequent evolution) along with the surface integral of the complex scalar field, $\Phi$. The black hole forms at $t\approx100$, producing a characteristic burst-type gravitational wave, observable at the corresponding retarded time of $t-r=100$. This gravitational wave closely resembles the merger and ringdown observed in head-on black hole collisions. Following the passage of the gravitational wave at $r=140$, a significant burst of coherent scalar radiation is seen to escape from the influence of the newly formed black hole.

In the previous compact configuration, a black hole was observed to form near the origin (as detected by the \texttt{AHFinder} thorn \cite{Thornburg:1995cp}) shortly after the chain’s components merged. It would be valuable to further investigate whether compact dipoles or higher-order chains could collapse simultaneously into multiple black holes that later merge, or under what conditions the Einstein-FLS theory might allow for Schwarzschild-like black holes balanced by scalar field interactions. This kind of equilibrium solutions have been constructed recently in the EKG theory with a sextic self-interacting scalar potential \cite{Herdeiro:2023mpt}.

Before concluding this section, we offer a final remark on the stability of gravitating chains. Preliminary results indicate that $k_z=3$ and $k_z=5$ configurations with $\mu = 0$ are unstable. These simulations are computationally two to three times more demanding than those for the dipole case, so a comprehensive exploration of the parameter space is deferred to future work. This future analysis will also include the construction and stability study of odd chains in the Einstein-FLS theory.

\section{Conclusion}\label{sec:conclusions}

In this work, we have demonstrated the existence of dipolar solutions within the Einstein-FLS model and explored their stability through full 3D nonlinear evolutions. Our results confirm the presence of stable configurations within specific regions of the parameter space. By scanning a range of parameter values, we have been able to identify configurations where the dipoles remain stable over long timescales even when additional forced perturbations are included, as well as cases where instabilities develop. The interplay between the gravitational attraction and scalar field interactions plays a key role in achieving equilibrium, but as we have seen, stability is not guaranteed across the entire parameter space.

We further demonstrated that for sufficiently small values of the real scalar field mass, $\mu$, non-rotating Q-chains (particularly dipoles and even-chains) can be constructed within the standard FLS model in flat spacetime. Comparable static configurations do not exist for a single scalar field with generic self-interactions, making two-component FLS configurations the simplest model of this type. While the dipoles are found to be robustly stable, Q-chains beyond the dipoles are found to be unstable, at least for the parameters explored here. These findings broaden the understanding of solitonic solutions beyond commonly studied monopolar configurations. By coupling four and six component flat chains to gravity, a sequence of gravitating chain solutions can also be constructed.

In the weak-field regime, a known formation mechanism for boson stars in the EKG model involves colliding configurations with opposite charges that subsequently start to swap while maintaining a quasistationary metric and energy density. It is likely that similar multipolar structures arising from the Einstein-FLS model can be realized, warranting investigation into their physical implications, particularly in cosmological and astrophysical contexts.


\acknowledgments
VJ acknowledges Carlos Herdeiro, Darío Núñez and Eugen Radu for their helpful suggestions and encouragement in the early phases of this research, which provided critical direction for the manuscript’s development. We would like to thank Guo-Dong Zhang, Qi-Xin Xie and Gabriel Luz Almeida for helpful discussions.
SYZ acknowledges support from the National Key R\&D Program of China under grant No.~2022YFC2204603 and from the National Natural Science Foundation of China under grant No.~12075233 and 12247103.

\bibliography{ref}


\end{document}